\documentclass[10 pt,journal,compsoc]{IEEEtran}
\ifCLASSINFOpdf
\else
\fi
\usepackage{multicol,graphicx,amssymb,mathrsfs, amsmath,pifont,amscd,latexsym,amsthm,fancyhdr,CJK,url,hyperref,longtable, pifont, subfigure, slashbox, colortbl, booktabs,balance}
\usepackage[nounderscore]{syntax}
\usepackage{caption}
\usepackage[linesnumbered,boxed,algosection]{algorithm2e}

\usepackage{listings}
\usepackage{color}

\definecolor{lightgray}{rgb}{.9,.9,.9}
\definecolor{darkgray}{rgb}{.4,.4,.4}
\definecolor{purple}{rgb}{0.65, 0.12, 0.82}
\definecolor{olivegreen}{rgb}{0.04,0.42,0.28}
\definecolor{orange}{rgb}{0.99,0.43,0.11}

\lstdefinelanguage{JavaScript}{
  keywords={typeof, new, true, false, catch, function, return, null, catch, switch, var, if, in, while, do, else, case, break, foreach, boolean, class, throw, this, datetime, export, throw, implements, import, where, when},
  keywordstyle=\color{blue}\bfseries,
  ndkeywords={ entity, operation, var, handler, screen, header, footer, body, widget, touch, timeclock},
  ndkeywordstyle=\color{olivegreen}\bfseries,
  identifierstyle=\color{black},
  sensitive=false,
  comment=[l]{//},
  morecomment=[s]{/*}{*/},
  commentstyle=\color{purple}\ttfamily,
  stringstyle=\color{black}\ttfamily,
  morestring=[b]',
  morestring=[b]"
}

\begin{document}

%
\title{MUIT: A Middleware for Adaptive Mobile Web-based User Interfaces in WS-BPEL}
\author{Xuazhe~Liu~\IEEEmembership{Member,~IEEE}, Mengwei~Xu,  Teng~Teng, Gang~Huang~\IEEEmembership{Member,~IEEE,} and Hong~Mei~\IEEEmembership{Fellow,~IEEE,}

\IEEEcompsocitemizethanks{\IEEEcompsocthanksitem Xuanzhe~Liu, Gang~Huang, Mengwei~Xu, and Hong~Mei are with the Key
Laboratory of High Confidence Software Technologies (Peking
University), Ministry of Education, and Institute of Software,
School of Electronics Engineering and Computer Science, Peking
University. Beijing, China, 100871. Email: \{liuxuanzhe, hg, mayun, xumengwei\}@pku.edu.cn. \newline
Teng~Teng is with the Kingdee International Software Group Company Ltd. Shenzhen, China. Email: teng$\_$teng@kingdee.com. \newline
\protect\\
Corresponding author: liuxuanzhe@pku.edu.cn
}
\thanks{Manuscript received December, 2015}}


%

\markboth{Technical Report,~Vol.~XX, No.~XX, XXXX~2015}%
{Liu \MakeLowercase{\textit{et al.}}: MUIT: A Middleware for Adaptive Mobile Web-based User Interfaces in WS-BPEL}


\IEEEcompsoctitleabstractindextext{%
\begin{abstract}
In enterprise organizations, the Bring-Your-Own-Device (BYOD) requirement has become prevalent as employees use their own mobile devices to process the workflow-oriented tasks. Consequently, it calls for approaches that can quickly develop and integrate mobile user interactions into existing business processes, and adapt to various contexts. However, designing, developing and deploying adaptive and mobile-oriented user interfaces for existing process engines are non-trivial, and require significant systematic efforts. To address this issue, we present a novel middleware-based approach, called MUIT, to developing and deploying the \textbf{M}obility, \textbf{U}ser \textbf{I}nteractions and \textbf{T}asks into WS-BPEL engines. MUIT can be seamlessly into WS-BPEL without intrusions of existing process instances. MUIT provides a Domain-Specific Language (DSL) that provides some intuitive APIs to support the declarative development of adaptive, mobile-oriented, and Web-based user interfaces in WS-BPEL. The DSL can significantly improve the development of user interactions by preventing arbitrarily mixed codes, and its runtime supports satisfactory user experiences. We implement a proof-of-concept prototype by integrating MUIT into the commodity WS-BPEL-based Apusic Platform, and evaluate the performance and usability of MUIT platform. 

\end{abstract}

}

\maketitle

\IEEEpeerreviewmaketitle

\section{Introduction}
\IEEEPARstart{I}{n} the past decade, WS-BPEL~\cite{IBM:WS-BPELV2} has been very popular in enterprise computing environment together with useful supporting tools. The initial design goal of WS-BPEL assumes that all activities are realized by the automated interactions among several Web services. In the process engines that follow WS-BPEL specification, a set of Web services are automatically executed for completing a long-running tasks, without any interruption to involve human interactions and tasks. However, in practice, people often participate in the execution of business processes, while requiring new aspects such as interactions between the process and the user interface. Hence, some complementary extensions such as BPEL4People~\cite{IBM:WS4People07} and WS-HumanTask~\cite{IBM:HumanTask07} were proposed.

\indent Since the announcement of Apple iPhone in 2007, the sale of mobile devices such as smartphones and tablet computers keeps fast growing. People are used to taking these mobile devices to process various daily tasks, such as making search, browsing news, processing E-mails, and so on. Besides personal purposes, mobile devices are also used in the enterprise computing. More and more companies adopt the so-called \textbf{BYOD} (\textbf{B}ring-\textbf{Y}our-\textbf{O}wn-\textbf{D}evices) policy in priority. Employees can take smartphones and tablet computers to process business tasks. There have been a large volume of  scenarios for using mobile devices in enterprise-level business processes. For example, an employee can initiate his reimbursement process, and share the scanned receipts via DropBox or Google Drive. The manager who is now on business travel can receive a notification on smartphones, check details, and submit a form for approval. 
 
\indent Although urgent requirements exist, developing and deploying mobile-oriented user interfaces into current business process are non-trivial tasks. From software engineering perspective, some technical issues need to be addressed.  

\noindent \textbf{Seamless integration}. Similar to BPEL4People and WS-HumanTask, we need extensions to specify user interface descriptions of the activities that should be performed on mobile devices. The descriptions should be seamlessly integrated into existing process description specifications and running instances. The descriptions should support standard data exchange with current workflow constructs,  and enable control flow decisions based on the data input from previous (automated or manual) activities. In addition, learning the lessons from the unsuccessful adoption of BPEL4People and WS-HumanTask~\cite{Kunze:JCP07}\cite{Russo:SPE12}, composing these mobile user interactions should not impact existing WS-BPEL processes and execution engines. 

\noindent \textbf{Rapid adaptive mobile UI development.} Designing and implementing user interfaces for mobile devices need to take into account some new features. First, employees' devices may have different OS platforms such as iOS and Android. Hence, a platform-neutral abstraction is preferred to simplify user interfaces development and deployment. Second, user interactions patterns on modern smartphones and tablets are quite different from desktop PCs, traditional feature phones, and PDAs. Expected user interaction features include limited screen estate, touch-centric control, push-oriented notification, context-aware responsiveness, adaptation to portrait or landscape, and so on~\cite{Hemel:OOPSLA11}. In practice, the user interfaces design is quite time-consuming and tedious, as it needs to cover a lot of aspects such as styling, layout, and so on. The preceding mobile-specific features increase the complexity of UI development, since they are likely to be casually mixed with application logics. Hence, it calls for a more \textit{well-structured, mobile-specific, and adaptive} user interface abstraction that can be fast and easily deployed into existing business processes. 

\noindent \textbf{Efficient communication performance.} When integrating mobile user interactions into WS-BPEL processes, mobile devices are connected to and interact with the process engine server. Due to human mobility, the network connection is not always available and reliable. Although the computation resources like CPU, GPU, RAM, and battery keep improving, it is yet impossible and impractical for mobile devices to continuously interact with the remote WS-BPEL engine server, or performing long-running computation-intensive tasks. Hence, the integration solution should take into account efficient communication between mobile devices and process server.

To address the preceding issues, we propose a novel middleware-based approach, called \textbf{\textbf{MUIT}}, which refers to supporting \textit{\textbf{M}}obile \textit{\textbf{U}}ser \textit{\textbf{I}}nteractions and \textit{\textbf{T}}asks in WS-BPEL. \textbf{MUIT} is realized as a standard Web service that can be seamlessly integrated into existing WS-BPEL engines. Middleware has been proved to be a promising approach to supporting mobile computing applications with the provision of a highly configurable and adaptive execution environment that dynamically reacts to changes in operating contexts~\cite{Chan:TSE03}. Conceptually, MUIT provides the process developers  simple syntax and open application programming interfaces (APIs) to implement the Web-based UI and, if required, to dynamically adapt the UI to respond to various changes in the contexts. 
The core of MUIT approach is a Web-based UI programming abstraction with its Domain-Specific Language (DSL). Developers can use our DSL to declaratively define rich and adaptive mobile-oriented Web UI by simple JavaScript-like syntax. We design a mechanism that automatically generates MUIT UI from standard WSDL descriptions. We have deployed MUIT on a commercial WS-BPEL product, called  Kingdee Apusic Platform Suite V6~\footnote{http://en.kingdee.com. Kingdee is a leading middleware and enterprise solution provider in China. One co-author in this paper serves as the director of Kingdee middleware research laboratory and we deployed a beta version of Apusic Platform with MUIT.}, and demonstrated our approach's usability and performance in real-world case scenarios. 

\indent The main contributions of MUIT can be summarized as follows.  
 \begin{itemize}
 \item We identify the key technical requirements for integrating mobile-specific user interfaces in terms of system integration, UI development, and performance.
  \item We present a non-intrusive design that can be seamlessly integrated into existing running business processes without additional extensions or modifications.
  
 \item We propose a programming abstraction together with its domain-specific language to help developers more efficiently develop adaptive Web UI compliant to MVC pattern. Based on some typical illustrating scenarios, we demonstrate that the programming complexity caused by mixture of codes (HTML, JavaScript, and CSS) is reduced and the UI generation is automated.  
  \item We provide efficient runtime mechanisms to optimize communication performance at both server side and client side. The mechanisms can significantly reduce the unnecessary network resource allocation, computation, and energy cost.
 \end{itemize}
 
The remainder of this paper is organized as follows. Section 2 presents a motivating example that illustrates mobile user interaction requirements and characteristics in WS-BPEL. Section 3 describes our approach overview by modelling interactions among MUIT, WS-BPEL and mobile users. Section 4 presents the design patterns of MUIT. Section 5 provides the details on how our DSL is designed and implemented in MUIT. Section 6 describes how MUIT optimizes runtime performance issues and evaluates the system usability. Section 7 compares some related work. Section presents some discussions and Section 9 concludes the paper with conclusion remarks.  
\section{Motivations and Requirements}
We begin with our approach by a motivating example and clarify the technical requirements of integrating mobile user interfaces in WS-BPEL. 
\subsection{Motivating Example}
We describe a simple reimbursement example which is commonly deployed in many corporations. Suppose that an employee wants to reimburse business travel expenses. He submits a request form to the reimburse WS-BPEL process. The reimbursement process sequentially executes four Web services: (1) the \texttt{IDAuthentication} service checks the employee's identity and department; (2) if the authentication is passed, the \texttt{TaskForwarding} service is invoked to forward this reimbursement request form to the employee's manager's task lists; (3) the \texttt{Notification} service is then triggered to send a notification to the manager's smartphone via SMS, Instant Messaging Services, or E-mail; (4) when clicking through the notification, the manager will be redirected to the \texttt{TaskApproval} service, from which he can view the details (e.g., electronic receipt proofs) of this reimbursement request and decide to approve or decline it. If the manager could not process it immediately, he can delay the task by setting it as a ``\textit{To-Do}" event on his calendar. The manager is possibly leaving for a meeting where network connections are unavailable, and he may process the task somewhere later, e.g., on taxi. The results should be cached on the smartphone and be synchronized back to the server when network connection recovers. When accomplished, the \texttt{TaskApproval} service will update the status of this reimbursement request and invoke \texttt{Notification} service to deliver processing results to the employee.
\subsection{Characterising Technical Issues}
Based on the preceding example, we analyze some key technical issues to realize the requirements of integrating user interfaces into WS-BPEL.
\subsubsection{System Integration Requirements}
In the example, the reimbursement process requires human intervention when invoking \texttt{TaskApproval} service. WS-BPEL was originally designed for automating a set of Web services, so the first requirement is \textit{how to identify} \texttt{TaskApproval} \textit{as a} \textbf{human-intervention} \textit{service} and \textit{how to integrate its user interfaces into existing process engines}. One possible solution is to use common extensions for user interfaces like BPEL4People or WS-HumanTask. However, it is known that these solutions mostly concentrate on the specification of manual human-oriented tasks, but lack the possibility to describe a user interface in detail~\cite{Kunze:JCP07}. In addition, although these solutions are built upon WS-BPEL specification, deploying them actually needs some modifications of existing WS-BPEL engines. Hence, in our opinion, we need to provide more practical solution of integrating UI integration into WS-BPEL process. In addition, the integration should be as non-intrusive as possible, and have few impacts on existing running infrastructures. 
\subsubsection{Adaptive UI Development Requirements}
The second requirement is to develop the user interfaces for \texttt{TaskApproval} service, such as basic information fields, forms, calendar, and other rich widgets. The first issue we need to consider the \textbf{device diversity} in employees, i.e., different OS platform vendors such as iOS, Android, BlackBerry, Windows Phone, and so on. Although they may have similar user interactions such as touch-centric control, the implementation techniques are quite different. For example, iOS uses Object-C and Android uses Java. If we choose the native apps for enabling \texttt{TaskApproval} service, we have to develop and maintain various versions of apps for different platforms. Therefore, a platform neutral design is preferred, e.g., the Web applications (a.k.a., Web apps). 

\indent Besides the platform neutrality requirement, in the example, some unique user interaction features should be also addressed in the UI design to adapt user operations.

\noindent $\bullet$ \textit{\textbf{Limited screen estate}}. Undoubtedly, the screen size of screen is often small, limited, and various. For example, the screen estate of smartphones usually ranges from 3-inch to 6-inch. Limited screen estates may have impacts on task information presentation and user inputs. Tasks are displayed with rather small font size, or users have to scroll down several times for viewing the complete task information. Some adaptations are required, such as font size, splitting tasks into multi sub-pages, and displaying them on several screens.

\noindent$\bullet$ \textit{\textbf{Responsive to changes}}. Due to the portability of devices, the task UI is expected to respond to context changes, such as holding the device in vertical or horizontal orientation, portrait or landscape, and changes in location.

\noindent$\bullet$ \textit{\textbf{Touch-centric controls}}. Compared to the mouse-and-keyboard centric operations on desktop PCs, expected user interaction patterns are a bit different on mobile devices. To process a task, people prefer simple touch controls and gestures such as tapping, swiping and pinching. 

\noindent$\bullet$ \textit{\textbf{Push-oriented notification}}. When a task is dispatched, it requires a mechanism to notify corresponding stakeholders. In some BPM specifications, it suggests maintaining a task list for corresponding roles, who are assumed to view and process tasks periodically. In other words, people are ``pulled" to perform the task processing. However, on mobile devices, the notification should be ``\textit{push-oriented}". A prompt window of SMS, E-mail, or other Instant Messaging Services notifications can facilitate users to process the task in time. 

\indent The preceding mobile-specific features need to be comprehensively addressed and carefully integrated in user interface. Although there have been some frameworks such as jQuery Mobile~\footnote{http://www.jquerymobile.com} and Sencha-Touch~\footnote{http://www.sencha.com/products/touch}, developers still need to do a lot of manual and tedious tasks such as data binding, event handling, layout and styling, and so on. More seriously, without well-structured implementation model, the implementations of these UI features are likely mixed with application logics, and the maintenance and adaptation are hard to perform. An efficient programming abstraction is required to better organize the modules and to boost the development efficiency.
\subsubsection{Performance Requirements}
Compared to previous PDAs, current smartphones are with more powerful computation resources. In the preceding example, however, it may still take the manager quite a long time to process the task. The manager is also likely to be interrupted by a meeting. The device is still connected to network and the energy keeps consuming. In addition, the network connections on mobile devices are not always available, stable, or reliable, as people keep moving around. Therefore, the processing results may not be correctly submitted caused by timeout or disconnection. The user interfaces should be aware of such context changes, and persist data locally. In our opinion, the \textbf{offline task handling} is a desired solution. People can still work out tasks when they are under unavailable network connections. When the network connections recover, the processing results can be synchronized back to process engine. As a result, our UI design needs to support offline task handling. Certainly, WS-BPEL engines should take charge of managing the interaction states of several concurrently connected devices. 
\section{MUIT System Architecture}
To address the preceding technical issues, we then propose MUIT, a middleware-based approach to systematically designing, implementing, and deploying mobile Web-based user interfaces into WS-BPEL processes. 

Technically, MUIT plays as a middle-box between the WS-BPEL server and the mobile devices. It provides a programming abstraction that supports meta programming~\cite{Sheard:sigplan02} by organizing all required mobile-specific features into a well-defined model. Based on a Domain-Specific Language, developers can declaratively develop MVC-based UI specification with very simple annotation syntax, to avoid arbitrary code mixture of HTML, JavaScript, and CSS. The MUIT's compiler can automatically generate HTML, JavaScript, CSS, and other necessary resource files for the target UI. In this way, from all current popular platforms such as iOS, Android, and BlackBerry, users can access the MUIT UI via Web browsers such as Chrome, FireFox, Safari, and so on. Externally, an MUIT UI is published as a standard Web service. In this way, MUIt can be seamlessly integrated into standard WS-BPEL instances without introducing any new extensions and additional intrusions. In other words, existing WS-BPEL infrastructures are not be instrumented at all. 
\begin{figure}
\centering
\begin{center}
\includegraphics[width=0.45\textwidth]{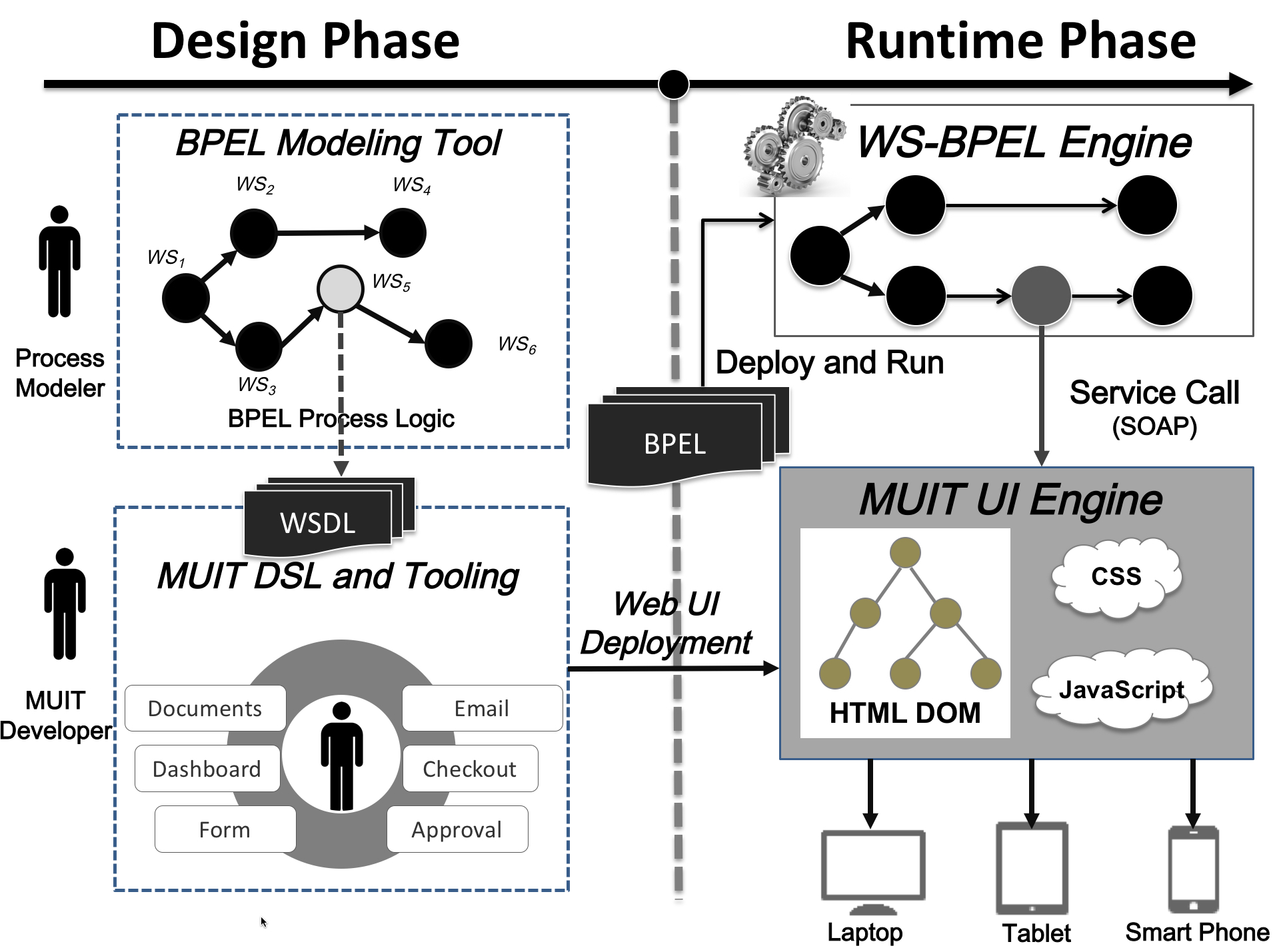}
\caption[7.5pt]{The MUIT Middleware}\label{approach}
\end{center}
\end{figure}

Essentially, our MUIT approach consists of two main phases, i.e., the \emph{\textbf{design phase}} and the \emph{\textbf{runtime phase}}, as shown in Figure~\ref{approach}.  

\indent At the design phase, the process modellers design and specify business process logics in standard WS-BPEL files and deploy them on WS-BPEL engines. Here, neither editors nor engines need additional supports for MUIT. When a Web service in the process requires human invention and user interfaces, the process modellers should specify it as an MUIT service with endpoint reference, and import the WSDL file to MUIT Development Tool. The tool generates an intermediate but quite simple UI according to the service interfaces, inputs, and outputs. The MUIT developers use the DSL to refine the task UI for supporting context-awareness. Some UI constructs for mobile tasks, e.g., notification, dashboard, form, approval, and checkout, are developed as built-in features. The MUIT service is deployed and published on the MUIT engine as a regular Web service. In this way, the MUIT UI can be directly and seamlessly bound into standard WS-BPEL infrastructures, e.g., specified by an $<$\emph{invoke}$>$ activity in WS-BPEL specification. 

\indent When an MUIT service binding is made, the runtime SOAP requests to MUIT service will be processed by the MUIT engine. The engine interprets the DSL to regular HTML files, and sends the files to the users' mobile devices. When users perform tasks, the WS-BPEL process instance needs to be suspended to wait for task completion on mobile devices. Learning lessons from our previous \textit{iMashup} platform~\cite{Liu:TSC14}, a real-time monitoring service implemented in JavaScript is dynamically injected into the mobile-side browser to keep connections with MUIT engine. When a SOAP request arrives, the MUIT engine will create a new pending service instance where a monitoring service will instantiate a corresponding task UI. When the task is completed, the monitoring service will notify the MUIT engine. Then, the engine sends the result as SOAP response to the WS-BPEL engine. At all runtime, WS-BPEL engine sends SOAP requests to the MUIT engine and receives standard SOAP responses. The MUIT engine delivers task UI to mobile devices and gets responses by means of standard HTTP and JSON (JavaScript Object Notation). Underlying details such as how to support synchronization or asynchronization among mobile devices, the MUIT engine and WS-BPEL engine, and how the Web UIs are generated and adapted, are to be described in \S 4 and \S 5. 

\section{Design Patterns in MUIT}
\begin{figure}
\centering
\begin{center}
\includegraphics[width=0.45\textwidth]{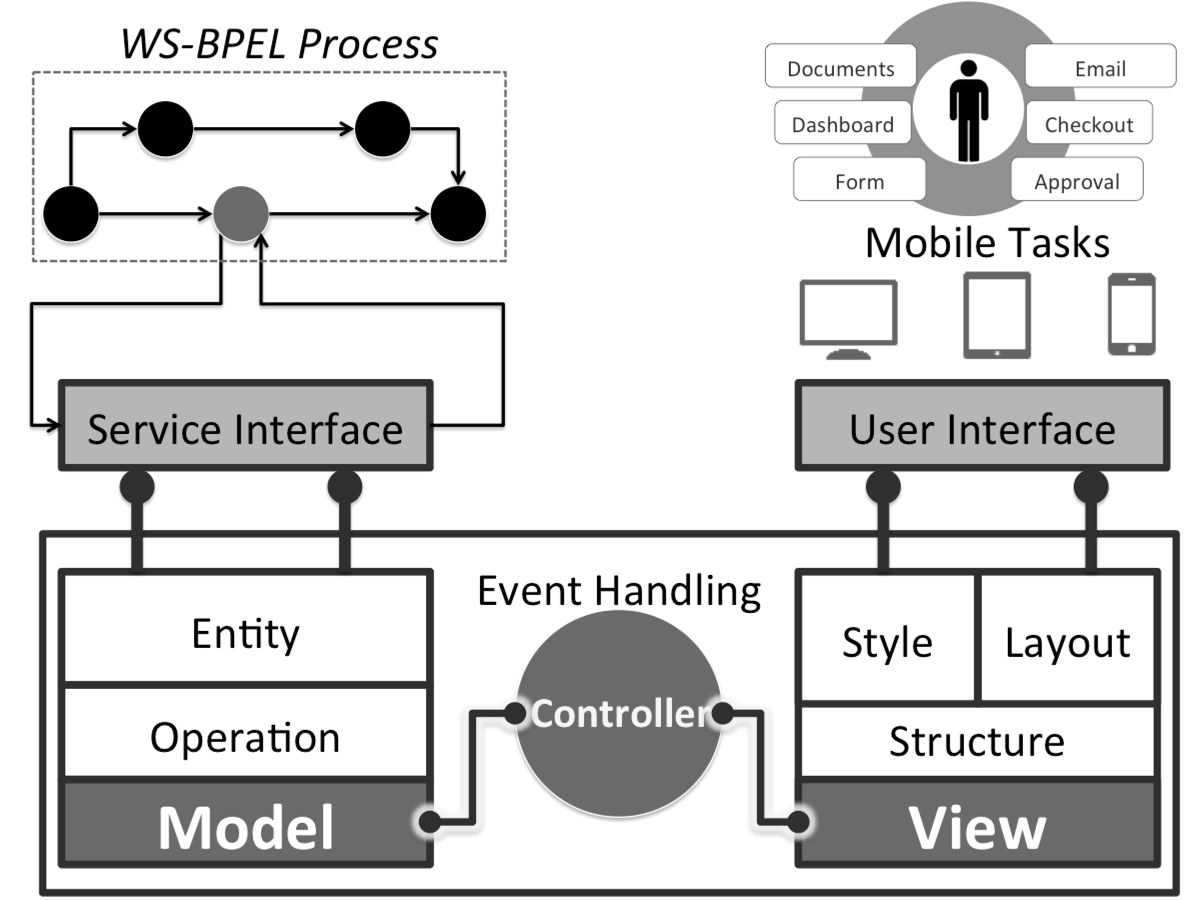}
\caption[7.5pt]{MUIT Service Model}\label{model}
\end{center}
\end{figure}

In this section, we first present the detailed design patterns of MUIT model based which the MUIT UI is implemented. To mediate the WS-BPEL process and the mobile users, the MUIT application model has two external interfaces, i.e., the \textit{Service Interface} and the \textit{User Interfaces}, as shown in Figure~\ref{model}. MUIT implements the communication with WS-BPEL process by the \textit{Service Interface}. The \textit{Service Interface} is described as a regular WSDL specification. Such an interface resolves the problem of seamless integrating MUIT with standard WS-BPEL. 

The \textit{User Interface} is realized by MUIT developers to implement task user interfaces accordingly. In practice, developing and deploying Web applications is quite time-consuming. A number of architectural patterns have been developed, among which the most commonly used one is classical \textit{Model-View-Controller} design pattern. MUIT leverages the MVC pattern as a guide line to develop and deploy mobile user interfaces. 
\subsection{MVC Design Patterns in MUIT}
Generally speaking, the MVC pattern defines some principles to develop interactive applications by separating the roles of \textit{Model}, \textit{View}, and \textit{Controller}. 
\begin{itemize}
\item \textbf{\textit{Model}} represents the data that can be fetched from Web services in original WS-BPEL process and manipulated by the application, e.g. task description, task type, and so on.
\item \textbf{\textit{View}} defines the user interface, presenting (elements of) the \textit{Model}, such as form, navigation, et al.
\item \textbf{\textit{Controller}} deals with user events and adapts the \textit{View} and \textit{Model} accordingly. 
\end{itemize}

Although most Web developers adopt the MVC pattern, it is reported that in practice the HTML, JavaScript, and CSS are often arbitrarily mixed without well-structured organization. In particular, the \textit{Controller} parts play the infrastructural role to coordinate \textit{Model} and \textit{View}, but the implementation often falls into a lot of similar codes that have to be manually written in the \textit{Controller}. Such a development style makes the Web applications with mixed codes of application logics and UI controls, and difficult to adapt the context changes in mobile-oriented workflows.

\indent MUIT differs from existing MVC frameworks by automating the \textit{Controller} between the \textit{Model} and the \textit{View}. In MUIT, the data defined in the \textit{Model} could be either local or remote, with some processing logics such as accessing database, Web service related APIs, or other asynchronous APIs. The \textit{View} defines the presentation structure of task user interface. The \textit{View} can be parameterized with one or more \textit{Model} objects sent by the \textit{Controller} to present. Additionally, MUIT provides some configurable parameters for specifying layout and styling. For example, front and background color, default size, initial location, and presentation mode (visible or hidden, maximized or minimized). As we will show later, developers can use the DSL defined by MUIT with a simple declarative syntax to implement the \textit{View}. 

\indent The \textit{Controller} automatically coordinates the \textit{Model} and the \textit{View}, i.e., responsible for instantiating the \textit{View} and populating them with data from the \textit{Model}. More specifically, the \textit{Controller} takes charge of communicating with Web services via \textit{Service Interface} and persisting fetched data to the \textit{Model}. To this end, the \textit{Controller} defines some functions to wrap the operations defined in the \textit{Service Interface}. Then, the \textit{Controller} reads data from the \textit{Model}, sends it to the \textit{View}, and manipulates the \textit{Model} according to user inputs on \textit{View} by means of \textit{event handling}. The events can be either provided by the underlying mechanisms (such as native HTML or JavaScript event models), or defined by the MUIT developers.
Usually, both user actions on the \textit{View} and requests from
other \textit{Views} may trigger events, and result in the request for the functionalities as well as the reaction of its own UI. We will provide more details on how MUIT automates the \textit{Controller} by its DSL in \S 5.5. 

\begin{figure}
\centering
\begin{center}
\includegraphics[width=0.45\textwidth]{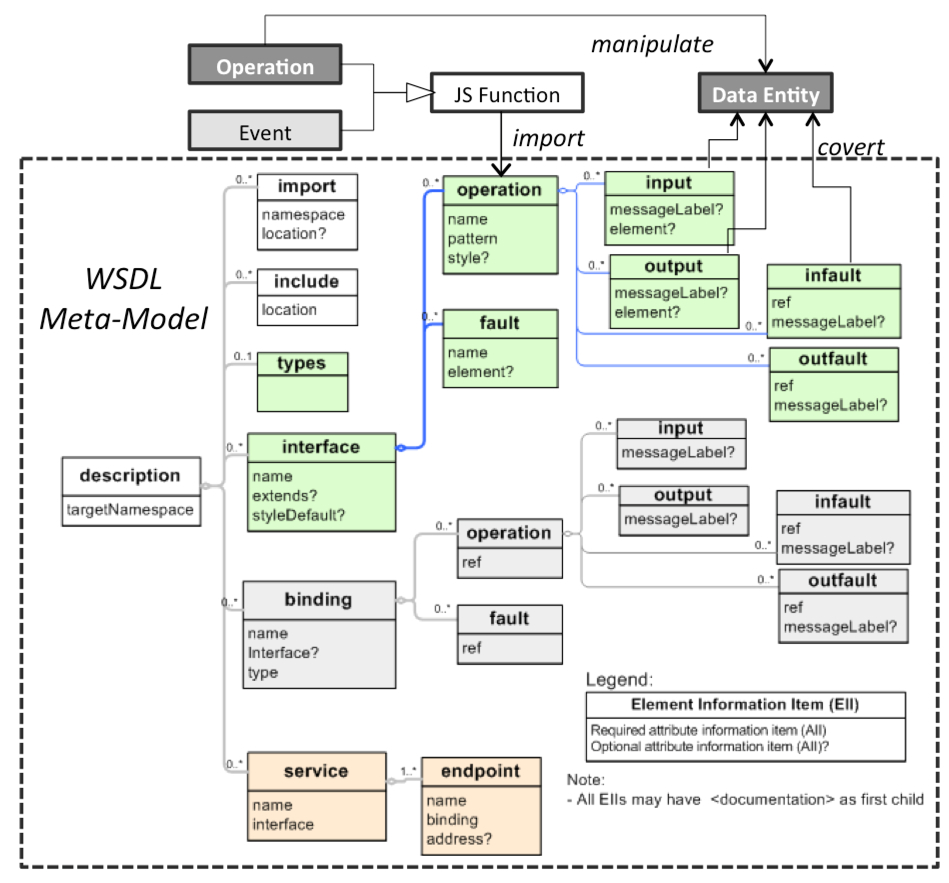}
\caption[7.5pt]{Model Transformation from WSDL to MUIT MVC }\label{transform}
\end{center}
\end{figure}
\subsection{Intermediate UI Automation}

We describe how MUIT realizes the communication with WS-BPEL process from \textit{Service Interfaces}. The \textit{Service Interfaces} is described as regular WSDL. The MUIT engine needs to transform the WSDL description into its own MVC pattern. Inspired by lessons~\cite{Florian:InfoSys12}and~\cite{Kassoff:IC03}, we realize the transformation from WSDL to the MUIT model as follows. 

\indent As illustrated in Figure~\ref{transform}, MUIT implements some processing logics over standard WSDL. We also show a sample of reimbursement task Web services and how the transformations are performed. By default, all \textit{input} and \textit{output} messages are extracted and stored as \textit{Data Entity} in the \textit{Model}. Entity names and types correspond to the messages type definition.  All operations defined in the \textit{WSDL Interface} are transformed as either \textit{Operations} performed over \textit{Data Entity}, or events defined in \textit{Controller}. Here, the port address of the defined service corresponds to the URL at which the actual web application html files can be downloaded for instantiation.

When receiving the requests of WS-BPEL process, the MUIT engine needs to transform the WSDL description into an intermediate MVC-compliant UI. For example, each variable in \textit{input} and \textit{output} messages are extracted and stored in the \textit{Model}. All operations defined in the WSDL \textit{Interface} are transformed as either some \textit{\textbf{operations}} in the \textit{Model}, or some events defined in the \textit{Controller}.


\indent Learning lessons from previous efforts~\cite{Florian:InfoSys12}\cite{Kassoff:IC03}, we provide some predefined templates to automatically generate default \textit{View} for a given task by parsing their \textit{Service Interface}, such as forms, tables, buttons, navigation, and so on. All input messages of a Web service operation will be assigned with basic HTML elements. For example, for the \textit{string}, \textit{number}, and any other elements, an HTML textfield ``$<$ \texttt{input type = "text"} $>$" will be generated. For fields which point to other data, a \textit{form} element will be created. In the \textit{Controller} definition, we need to associate the \textit{form} with \textit{action}, which uses the data, calls the method and updates the \textit{View}. Such a raw UI can be further refined by the MUIT developers with advanced UI features.

\begin{figure}
\begin{lstlisting}[numberstyle=\tiny]%

<wsdl:definitions name="reimbursementTask" targetNamespace="http://www.pku.edu.cn/muit/>
<!-- types definition -->
......
<!-- massages definition -->
<wsdl:portType name="reimbursementTaskPortType">
<wsdl:operation name="getTaskInfo">
.....//Other operations mapping Task data model 
</wsdl:portType>
<wsdl:binding name="reimbursementTask" type="tns:reimburementTaskPortType">
......
<soap:address location="http://www.pku.edu.cn/MUIT/reimbursementTask.js" />
......
  </wsdl:definitions>
\end{lstlisting}
\caption[7.5pt]{Sample WSDL: Task and Binding to MUIT}\label{wsdl}
\end{figure}

\section{Programming Abstractions of MUIT}
Although MUIT can automatically create a Web UI, two key technical problems remain unresolved. On the one hand, realizing the preceding mobile-specific features need programmers' manual efforts, i,e.,  carefully writing HTML, JavaScript, and CSS codes for each feature and gluing them together. On the other hand, developing the \textit{Controller} still requires a lot of tedious and time-consuming efforts such as data binding, UI refreshing, and so on. To reduce the complexity of programming and make the final UI more compliant to MVC, we then introduce our DSL that is designed for reducing programming complexity as well as improving UI automation. 

Before the detailed design our DSL, we first describe some background of DSL.
\subsection{DSL: In a Nutshell}
In the research communities of software engineering and programming language, Domain-Specific Language (DSL) is an important and efficient way to reduce the programming complexity as well as improve the productivity~\cite{Fowler:DSL11}. Essentially, DSL aims to realize the concept of meta programming~\cite{Sheard:sigplan02}, which makes a program designed to read, generate, analyse or transform other programs, and even modify itself while running. A DSL is a ``mini" language built on top of a hosting language (such as C, Java, and JavaScript) that provides a common syntax and semantics to represent concepts and behaviors in a particular domain. As summarized in~\cite{Maximilien:ICSOC07}, using or designing a DSL generally helps achieving the following goals.
\begin{itemize}
\item \textbf{\textit{High-Level Constructs}} by abstracting the programming task at a level higher than what is available with the host programming language constructs or its libraries. A DSL allows the domain concepts, actions, and behaviors to be represented directly in the new syntax.
\item \textbf{\textit{Terse Code}} as an effect of programming in a higher-level of abstraction. 
\item \textbf{\textit{Simple and Natural Syntax}} leads to easy to write and read code. 
\item \textbf{\textit{Ease of Programming}}, which is desirable of any programming language and also somewhat difficult to judge. However, since a DSL enables the expression of constructs that map directly to a domain, it generally makes programming easier (for applications in the domain) than using the underlying language directly.
\item \textbf{Code Generation} is how a DSL primarily functions. Essentially, the DSL statements are translated at runtime into code that uses the underlying language and its libraries. This can be either using meta-programming techniques or by code generation of program files.
\end{itemize}

\subsection{DSL Constructs in MUIT}

In practice, there have been a lot of DSLs proposed for various contexts for Web applications, such as \textit{WebDSL}~\cite{visser2008:webdsl}, \textit{mobl}~\cite{Hemel:OOPSLA11}, \textit{MobiDSL}~\cite{Kejriwal:OOPSLA09}, and Web mashup DSL~\cite{Maximilien:ICSOC07}. 

The DSL of MUIT not only learns the general design principles of these DSLs, but also advances itself to support mobile workflow contexts. Our DSL supports the MVC design pattern by defining the Web-based user interfaces with some syntactic \textit{\textbf{constructs}}. Our DSL is designed for accommodating the mobile features that are required for user interactions of WS-BPEL. The DSL of MUIT defines some simple JavaScript-like syntax to help the MUIT developers specify the programming abstraction of \textit{Model}, \textit{View}, and \textit{Controller} into a well-defined structure that adapts the Web user interface in mobile workflow. Web elements of the intermediate UI are automatically described by our abstraction so that MUIT developer can declaratively refine the UI themselves. The MUIT engine takes charge of interpreting the UI abstraction and generating the HTML, JavaScript, and CSS for the final Web pages. 

We show our DSL's syntax that is described in the form of BNF (Backus-Naur Form), as shown in Figure~\ref{syntax}. We define some concepts such as \textit{entity}, \textit{operation}, \textit{screen}, and so on. The expressions, statements, and operators can be realized by developers. In addition, MUIT provide some simple APIs that can help developers deal with some common logics. For example, the API \texttt{import} can read the MUIT service's WSDL document from its \textit{URL} and passes the XML into JSON format. When a specific \textit{context} is triggered, e.g., holding the device in portrait or landscape mode, the UI needs to be changed by a corresponding \textit{adaptation} actions controlled by some \textit{rules}. 

The meanings of the DSL's syntax are mostly straightforward and easy to understand. We then move to describe the details of how our DSL realizes the MVC-compliant adaptive Web UI. In most of the following cases, we will demonstrate the UI that is developed by the MUIT DSL.
\begin{figure}[h]
\begin{grammar}

<definition> :: = ''\textit{entity}''| ''\textit{operation}''|''\textit{handler}''\\
 |''\textit{screen}''|''\textit{widget}''|''\textit{touch}''
 
<statement> ::= <var> ''='' <exp> 
\alt ''foreach'' ''(<exp>'')'' ''\{''<statement>*''\}'' 
\alt ''return'' <exp>? '';''
\alt ...

 <exp> ::= ''\textit{string}''|''\textit{int}''|''\textit{DateTime}''|
 \alt <exp> <op> <exp>
 \alt ''\{''<statement>*''\}'' 
 \alt ...
 
<op> ::= ''+''|''-''| ''*''| ''\%''|
\alt ... 
 
<context> ::= <when> ''('' <exp>  '')''
\alt <where> ''('' <exp>  '')''

<rule> ::= \textit{if} ''(''<context>'')'' {<adaptation>} 
\alt \textit{elseif} ''(''<context>'')'' {<adaptation>} *  
\alt \textit{else} ''(''<context>'')'' {<adaptation>} ?  

<adaptation> ::= ''screen'' ''<'' HTMLTag* ''>'' ''\{'' <function>* ''\}''''<'' /HTMLTag* ''>''
\alt ...

<api> ::= \textit{import}|\textit{exist}|\textit{navigate}... 

\end{grammar}

\caption{Syntax of MUIT DSL}\label{syntax}
\end{figure}

\subsection{Data Model Abstraction}
As shown in the MVC design pattern, the \textit{Model} represents the data fetched from Web services in original WS-BPEL process and to be manipulated by the application. However, processing the various types and formats of data from different Web services needs tedious programming efforts in traditional Web application developments~\cite{Maximilien:ICSOC07}. Hence, we define in our DSL with a \textit{data model abstraction} to enable the declarative style of realizing data models.

In MUIT DSL, the \textit{Data Model} consists of \textbf{entities} and \textbf{operations}. 

\noindent $\bullet$ \textbf{Entity}. Each \textit{\textbf{entity}} corresponds to an item from the Web service operation, e.g., task name, task description, and stakeholders. Each entity has a unique ID, a name, a set of properties, and associated functions defined by the \textit{Controller}. Each property has a name, a type, and optional annotations. 

\noindent $\bullet$ \textbf{Operation}. Each \textbf{\textit{operation}} contains the JavaScript APIs that invoke remote Web services or access local database. Typically, an \textit{\textbf{operation}} includes a list of input parameters accepting pass-in values and return values. In practice, these JavaScript APIs can be either synchronous or asynchronous. Asynchronous invocations are passed as a \texttt{callback} function and returned immediately.  

\indent Our DSL defines the data model abstraction for dealing with the data model from the MVC pattern. To demonstrate how data modelling works by MUIT DSL, we describe an example of an approval task in the reimbursement process. 

The sample code in Figure~\ref{approval} realizes the definition of the task UI with MUIT. The task consists of two entities: the \textit{Task} and the \textit{Role}. The \textit{Task} is specified by a name, a description, a status to indicate whether it is approved or is waiting for approval, a creation date, a due date, a role that is responsible for processing the task. A \textit{role} is a collection of people, including initiator, dispatcher, coordinator, approver, manager, and so on. The operation \texttt{import()} is mandatory for all MUIT services. Such an operation invokes the underlying API that reads MUIT WSDL document from its \textit{URL} with user name and password authentication.

Developers can define the data model with code fragments shown in Figure~\ref{approval}. The operation \texttt{getTaskInfo()} converts the data objects into a Task object \texttt{t} by the method \texttt{Task.fromTaskList()}. When the approval task is completed, the operation \texttt{approveTask()} updates the status and submits result back to WS-BPEL engine. The function \texttt{deplayTask} defined on \textit{Task} means that a task could be postponed for sometime, i.e., by modifying the due date. 
\begin{figure}
\begin{lstlisting}[numbers=none, numberstyle=\tiny]%

entity Task
	{
		String task_name:  Employee Travel Fee Approval;
		boolean status: waiting for approval;
		DateTime createDate: 2014-07-21;
		DateTime dueDate: 2014-07-22;
		String task_review;
		role:  Role<manager>;
		tags: reimbursement, travel, hotel, taxi;		
	}	
entity Role
	{
		String role: initiator | coordinator | dispatcher| reviewer ;
		task: approve<Task>;		
	}
operation import (String WSDLUrl, String user, String pwd)
	{
	 	var taskList =  httpRequest ("/WSDLUrl?user=" + user + "&pwd="+pwd); 			 	
	}
	
operation getTaskInfo()
{
	foreach (t in taskList)
	 	{
	 		add(Task.fromTaskList(t));
	 	}
}
operation approveTask(Task t)
{
	t.status = "approved";
}
operation delayTask(Task t, int days, String reason)
{
	t.status = "delay"; 	
	t.dueDate = DateTime.create
	(t.dueDate.getYear() + t.dueDate.getMonth()+t.dueDate.getDate() + days);
	t. reason = reason;
}
operation searchTask(tasktList, String s)
{
	if (s in taskList)
		return s;
		return false;
}

\end{lstlisting}
\caption[7.5pt]{Approval Task Data Model}\label{approval}
\end{figure}

\indent The data model is semi-automatically generated from the MUIT Web service's definition (by the operation \textit{import()}). In an ideal case, each element in WSDL can correspond to an \textbf{\textit{entity}}, and each interface can be mapped to an \textbf{\textit{operation}}. We allow the MUIT UI developers to refine the data model.  As elements in WSDL are structurally defined in XML, which is not efficient in Web-based user interfaces, we convert them to lightweight JSON objects. As we will shown later, such a design can significantly reduce the runtime overhead of MUIT. 
\subsection{User Interface Abstraction}
The user interfaces of MUIT is actually a HTML page that displays the data in DOM structure and provides the user interaction parts such as buttons and widgets. Generally, HTML defines the structure of a Web page by DOM tree, while CSS is employed to decorate the styling, like fonts, color, background, by applying CSS selectors. 

Native HTML and CSS do not support rich user interactions, such as calendar, time clock widgets, and other touch-centric features. Other popular JavaScript frameworks such as jQuery Mobile~\footnote{http://www.jquerymobile.com} and Sencha-Touch~\footnote{http://www.sencha.com/products/touch} provide a lot of pre-defined and built-in widgets, and can dynamically adapt them to the DOM. However, these widgets are framework specific. Developing new user-defined controls and widgets, such as the notification, page navigation, and adaptive display, is still non-trivial for developers. One possible way is to define some tags to annotate the HTML elements. However, it is tedious and efficient in practice. Hence, beyond library-level frameworks, our DSL provides higher-level abstraction to address the mobile-oriented UI features for WS-BPEL, so that developers can declaratively define the adaptive user interfaces by some core syntax constructs. We demonstrate the UI abstraction still by the task processing example.
\subsubsection{Page-Screen Presentation Unit}
A task UI is displayed on the screen of a physical smartphone or tablet computer. In other words, a screen is the basic display unit of a web page. Therefore, we introduce a construct called ``\textbf{screen}" to maintain a \emph{$<$ page, screen$>$} pair unit. A \textit{screen} actually refers to a Web page, consisting of HTML-compliant elements, variables, expressions, and contextual conditions. In MUIT, we default assign each task a home screen, just like the home page of a Web site. 

\indent A page screen construct encapsulates some useful JavaScript functionalities to get the information of the device model (e.g., \texttt{Samsung Galaxy Note 4}, \texttt{iPhone 4s}, etc.), screen size, and resolution level. In addition, it needs to detect the current orientation status by accessing the sensor information. Low-level JavaScript codes are not visible to the MUIT developers. Instead, these codes are automatically generated when deploying the UI on the MUIT engine. By such an abstraction, a lot of underlying trivial issues are hidden. MUIT allows some advanced adaptation expressions by two contextual conditions, \textit{where} and \textit{when}, which correspond to the \textit{context} construct in Figure~\ref{syntax}. For example, in the fragment of code shown in Figure~\ref{condition}, we declare a screen of approval task UI, and apply a conditional statement checking the device OS information. When iOS is detected, it requires to explicitly create a button for backward in the header. Such a policy is unnecessary to Android devices, as backward operation can be controlled by a physical button.
\subsubsection{Push-Oriented Notification}
Considering user's mobility,the MUIT server needs to push notification to the corresponding stakeholders instead of enforcing them to check on their own initiatives. Notification could be sent by SMS, E-mail, or other Instant Messaging Services. The notification usually includes the task name, followed by a hyperlink which can redirect to the task UI deployed on the MUIT engine. When receiving notification alerts, users can click through the hyperlink and navigate to the task UI within mobile Web browsers.
\begin{figure}
\begin{lstlisting}[numbers=none, numberstyle=\tiny]%

screen approveTask
{
	when (screen.deviceos == "iOS") 
	header (approveTask){	
	button {"back", history.go (-1);}}
}
\end{lstlisting}
\caption[7.5pt]{Adaptation with Device Type}\label{condition}
\end{figure}

\subsubsection{Tree-based Navigation among Screens}
Processing a task in workflow usually requires several steps. Take the reimbursement review task for example. The manager needs to view the task list, select and review a task item, and approve or delay the task. To enable the user interactions in MUIT, we need multiple screens and navigations between them. 

\indent On desktop PCs, navigation is triggered by clicking hyperlinks on the Web page, and redirects users from one page to another. Such a navigation is relatively casual and not strictly organized. On mobile devices, \textbf{the navigation between screens is a bit different and needs careful organization}. A typical organization is the tree-based, where each tree node can be realized as a single page-screen unit. It facilitates users to quickly jump from one screen to another. If a page contains deeper level information, a usual pattern is popping up a``\textit{cascading menu screen}". When the user selects an item from the list on the cascading menu, the current screen may slide to the left, and a new screen slides from the right. Hence, on mobile devices, navigation between screens may be moving forward into deeper level by using cascading menu, or moving back to a higher level by backward operation. However, the navigation among screen is not always strictly limited. We will illustrate how such an adaptation is realized in later sections.

\indent The tree-based navigation can be implemented by a stack of screens, where only the top of the stack is visible. Obviously, the top screen in the stack is the currently active physical screen. When an item is selected, a new screen representing the item is pushed onto the stack. When the backward operation is triggered, the current screen at the top is popped off the stack, and the previous screen appears again. However, on different platforms, the backward operation implementation is different. On iOS devices,  backward operation is realized by a \textit{back} button, which needs to be manually programmed by developers. In contrast, on most Android devices, the backward operation can be simply realized by pressing the physical button. As we described previously, the navigation pattern can be dynamically generated by detecting the OS platform in our MUIT DSL, with the condition constructs of \texttt{when} and \texttt{where}. In this way, implementing the stack-based navigation among screens is quite simple by means of function call stacks.
\begin{figure}
\centering \subfigure[iOS\label{decios}]{
\includegraphics[width=0.23\textwidth, height = 2.4in]{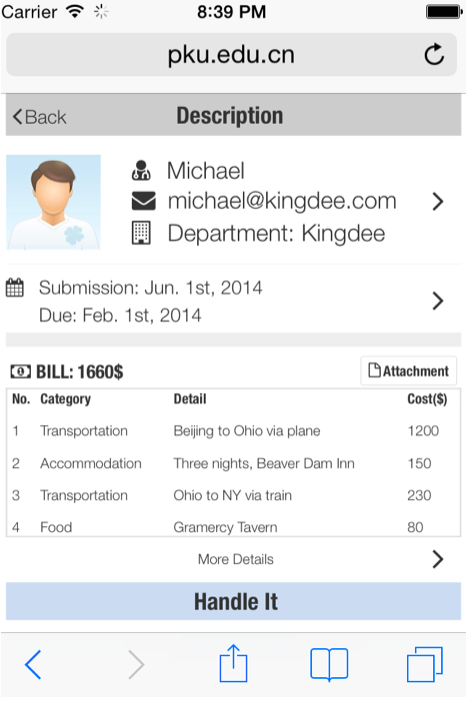}
}
\subfigure[Android\label{decandroid}]{
\includegraphics[width=0.23\textwidth, height = 2.4in]{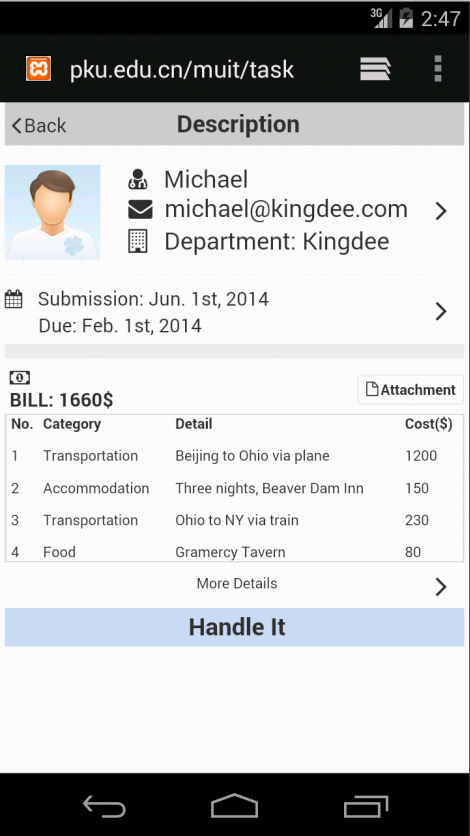}}
\caption[7.5pt]{Adaptive Styling on iOS and Android}\label{twoos}
\end{figure}

\subsubsection{User-Defined Touch-Centric Features}

\begin{figure}
\begin{lstlisting}[numbers=none, numberstyle=\tiny]%

var taskname = "Michael's US Travel Reimbursement"
var delaytime = "0";
var reason = "We delay... "
widget calendar c1()
{ 
	delaytime = select (option.value);
}
widget textInput tx1(String s)
{
	<input type = "text", value=reason/>
}
.......	
screen delayTask 
{
	header ("Delay");
	import(c1); 	
	import(tx1);
 	handler {
 	button { "Done", 
 				onClick = {delayTask(taskname, c1.delaytime, tx1.reason));}	}
 }
\end{lstlisting}
\centering \subfigure[iOS\label{delayios}]{
\includegraphics[width=0.23\textwidth, height= 2.4in]{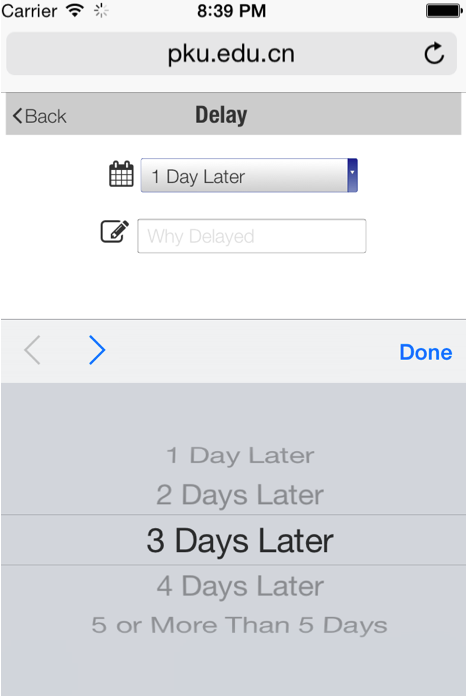}}
\subfigure[Android\label{delayandroid}]{
\includegraphics[width=0.23\textwidth, height= 2.4in]{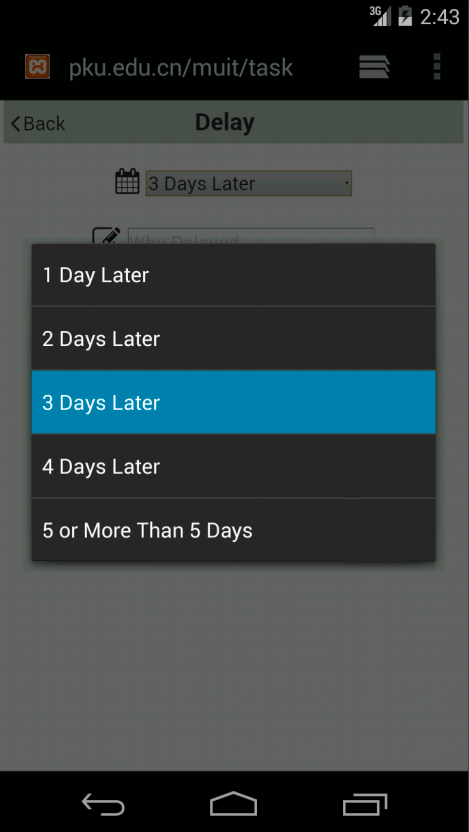}}
\caption[7.5pt]{ Delay Task by Advanced Controls}\label{delaycontrol}
\end{figure}
The preceding examples show only the basic HTML elements supported by MUIT. Actually, in MUIT, we support some advanced touch-centric features, beyond simple HTML elements such as \textit{text}, \textit{button}, \textit{checkbox}, and so on. These features mainly include two types: (1) \textit{widgets} such as weather, calendar, maps, and so on; (2) \textit{touch-centric controls} such as swiping, zooming, pinching, pressing, and so on. To this end, MUIT designs the constructs of \textbf{widget} and \textbf{touch} to help developers declare these interactions. 

\indent We take the delay task UI as an example. Figure~\ref{delaycontrol} demonstrates how the \textbf{widget} construct is used. We define a widget \texttt{calendar c1} so that the manager can extend the task due date by adding a \texttt{var} delaytime, whose initial value is 0. In addition, another widget \texttt{textInput} is defined to input the delay reason. Here, the \textit{calendar} and \textit{textInput} the are the types of \textbf{widget}. Furthermore, the \texttt{c1} can assign delay time by returning the \texttt{opion.value}, which is a regular HTML selector. The screen ``\texttt{delayTask}" is named with the \textit{header}, and uses \textit{import} to load \texttt{c1} and \texttt{tx1}. When the manager selects the delaytime, writes the comments for delay reason,  and presses ``Done" button, a handler is triggered to invoke the operation ``\texttt{delayTask(taskname, days, reason)}".

\indent In Figure~\ref{delaycontrol}, we show the different \textit{View} widgets on calendar on iOS (in  Figure~\ref{delayios}) and on Android (in Figure~\ref{delayandroid}), respectively. In our current implementation, we have some pre-encapsulated common widget libraries and touch events from jQuery mobile.

\indent Similar to \textbf{widget}, the \textbf{touch} construct can be attached to a screen to enable finger touch gestures. Considering the backward operation in Figure~\ref{condition}. If we don't want to use the ``\texttt{back}" button but swiping the screen from left to right for moving back to previous screen, we can define a \textit{touch} construct \texttt{ swipelefttoright()} in Figure~\ref{touch}.
\begin{figure}
\begin{lstlisting}[numbers=none, numberstyle=\tiny]%

touch swipe swipelefttoright (screen)
{
	var touchsurface;	
	function swipeDetect(touchsurface, callback)
	{	
		// logics to detect touch operation, start point and end point
	}
	history.back(-1);
}
screen approveTask
{
	header (approveTask){	
	import (swipelefttoright(approveTask));}}
}
\end{lstlisting}
\caption[7.5pt]{Touch Operation on Screen}\label{touch}
\end{figure} 
\subsubsection{Adaptive Display to Context Changes}
MUIT needs to adapt different interaction contexts to facilitate the user operations on mobile devices. In our current DSL, developers can declare the adaptive UI for three typical scenarios. 

\noindent$\bullet$ \textbf{Adaptive to Platform.} It refers to the adaptation on different mobile platforms such as iOS and Android. In Figure~\ref{delayios} and Figure~\ref{delayandroid}, we have shown the different UI displays on iOS and Android, respectively. Such adaptation is realized by a core function of our DSL, called \texttt{screen.devicetype()}. MUIT will dynamically load the corresponding CSS and JavaScript codes to decorate the user interface. 
\begin{figure}
\centering
\begin{center}
\begin{lstlisting}[numbers=none, numberstyle=\tiny]%

screen cascadingScreen(task t)
{
	header(t.taskName);
	t.getTaskInfo();
	button ("Back", history.go(-1);)
}
screen approveTask
{
	header ("TaskList");
	when ( (screen. window.innerWidth> 500) || 
	(screen.device.orientation == "horizontal"))
	{
		for each (t in TaskList)
		{
			t(onClick = {new cascadingScreen (t);}
		}
	}
}
\end{lstlisting}
\includegraphics[width=0.48\textwidth]{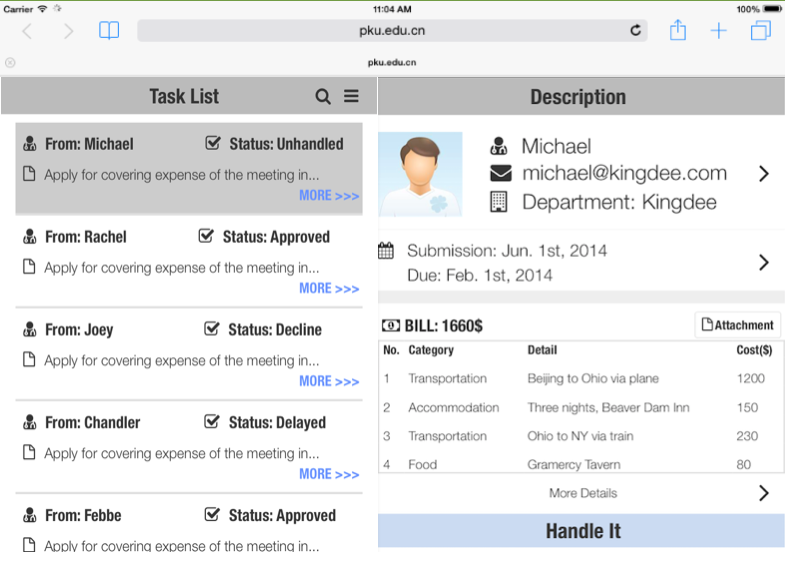}
\caption[7.5pt]{Adaptive Display with Screen Estate}\label{ipad}
\end{center}
\end{figure}

\noindent $\bullet$ \textbf{Adaptive to Screen Estate}. The second adaptive UI support is for screen estate. The screen size of mobile device  may vary significantly. On the narrow screens such as iPhone 4, a list of items is initially displayed. After an item is selected, its details will be put on a separate screen, just as the screen navigation mentioned previously. Then we can use the ``virtual" button on iOS or  physical back button on Android to move backward. In contrast, on larger screens, such as tablet devices or the smartphones with horizontal orientation,  a very common display pattern is ``cascading menu", which lists the detailed items by prompting up a new screen floating above the current screen. For example, when users browse on iPad as shown in Figure~\ref{ipad} or horizontal orientation (judged by the statement ``( \texttt{when ((screen. window.innerWidth $>$ 500) $||$ (screen.device.orientation $==$ "horizontal")})"), and click a task of``\texttt{Michael's reimbursement task}", MUIT initializes a new \texttt{cascadingScreen} (which instantiates the type of \textit{screen}) and lists the task information.

\begin{figure}
\centering
\begin{center}
\includegraphics[width=0.2\textwidth]{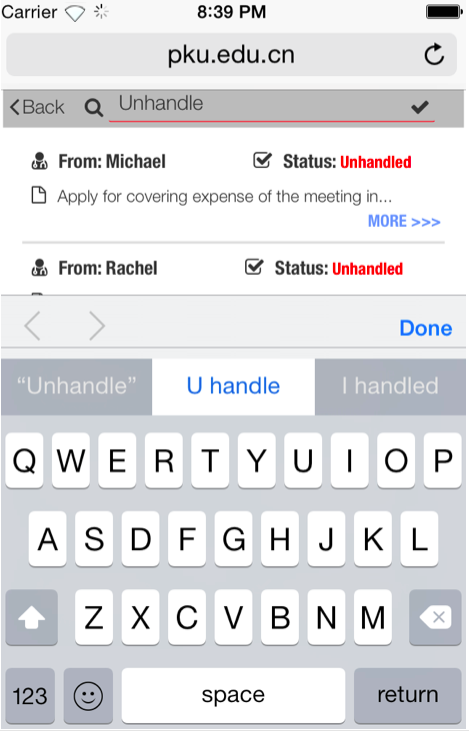}
\caption[7.5pt]{ Search from Local Storage Web Cache }\label{search}
\end{center}
\end{figure}

\noindent $\bullet$ \textbf{Adaptive to Offline Processing.}  We then introduce the third adaptation that deals with the possible network connection exceptions on mobile devices. Such a support is motivated by the fact that mobile users may loose stable connections (especially under cellular network) to the MUIT engine server and the WS-BPEL engine, but the task processing results are not submitted yet. To this end, MUIT leverages the HTML5 \texttt{appcache()} API~\footnote{\url{http://www.w3.org/html/wg/drafts/html/master/browsers.htm\#appcache}} to store all application codes and state on device local storage. The MUIT developers can optionally declare whether a task UI can be cached locally or not. If configured as ``true", all web elements of this UI are stored on device local storage (e.g., SD memory card) after they are downloaded from MUIT server for the first time. Application cache in MUIT stores data to local database by means of integrating SQLite. All intermediate data and state are cached. As shown in Figure~\ref{search}, even no network connections are available (note the network signal icon at the left top of the screen), MUIT users can yet search task list by phrase ``\textit{unhandled}" and get matched items. When network connection recovers, all data and state can be synchronized automatically.
\indent 

\subsection{Automating Controller by Abstraction}
In typical MVC design pattern, a change on the \textit{Model} will lead to changes on the \textit{View}. For example, when a new task item is added to the \textit{Model}, the \textit{View} will be updated automatically to show the information. Meanwhile, the \textit{Model} is updated when its properties are changed on the \textit{View}, e.g., by user inputs on a text field. In the \textit{MVC} design pattern, the \textit{Controller} takes charge of coordinating the \textit{Model} and the \textit{View}. 

However, in practice, implementing the \textit{Controller} needs several non-trivial tasks, such as requesting the data from the \textit{Model} (i.e., Web services) and writing back to the \textit{View}, manipulating the the \textit{Model} from the user inputs, and refreshing the \textit{View}. The \textit{Controller} plays an infrastructural role in MVC pattern, however, developers should write the JavaScript codes for each \textit{Controller} for every single coordination between the \textit{Model} and the \textit{View}. Leaving these imperative codes to developers may be quite tedious and inefficient, as developers have to rewrite these codes in many places with little or no alteration. Meanwhile, these codes for the \textit{Controller} is quite verbose. Hence, most of the \textit{Controllers} realize common functionalities to coordinate the \textit{Model} and the \textit{View}, we provide the abstraction, called \textit{event handler}, to automate the \textit{Controller}. 

\indent We still take the \textit{approveTask} operation as an example, as shown in the fragment of user code in Figure~\ref{controller}. MUIT generates a button with the value ``\texttt{approve}" for operation \texttt{approveTask()}. A variable \textit{taskname} represents the task name. Then, developers can declare an event handler as \texttt{onclick=approveTask(taskname)}" as the \textit{Controller} for associating ``\textit{approve}" button over the \texttt{approveTask} operation. In this way, the action on button widget is directly bound to the operation \texttt{approveTask(taskname)}. When a user switches to another task, the value of variable \textit{taskname} is automatically adapted.
\begin{figure}
\begin{lstlisting}[numbers=none, numberstyle=\tiny]%

var taskname = "Michael's US Travel Reimbursement"
<input type = "button" value = "approve"/>
.......	
 handler {
 	button { "approve", 
 				onClick = {approveTask(taskname));}	}
\end{lstlisting}
\centering
\caption[7.5pt]{Approve Task UI}\label{controller}
\end{figure} 

The abstraction of \textit{event handler} reduces the efforts of writing a lot of verbose codes for common functionalities in the \textit{Controller}. The \textit{event handler} directly associates the value or an attribute of a DOM node with an entity in the \textit{Data Model}. Learning lessons from our previous efforts on rich clients~\cite{Zhao:ICWS10}, the MUIT runtime provides an event bus through which each \textit{event handler} can subscribe some DOM nodes to monitor. When changes occur, the \textit{event handler} is triggered to update the \textit{Model} and the \textit{View} accordingly.
\section{Prototype and Evaluation}
\begin{figure}
\centering
\begin{center}
\includegraphics[width=0.45\textwidth]{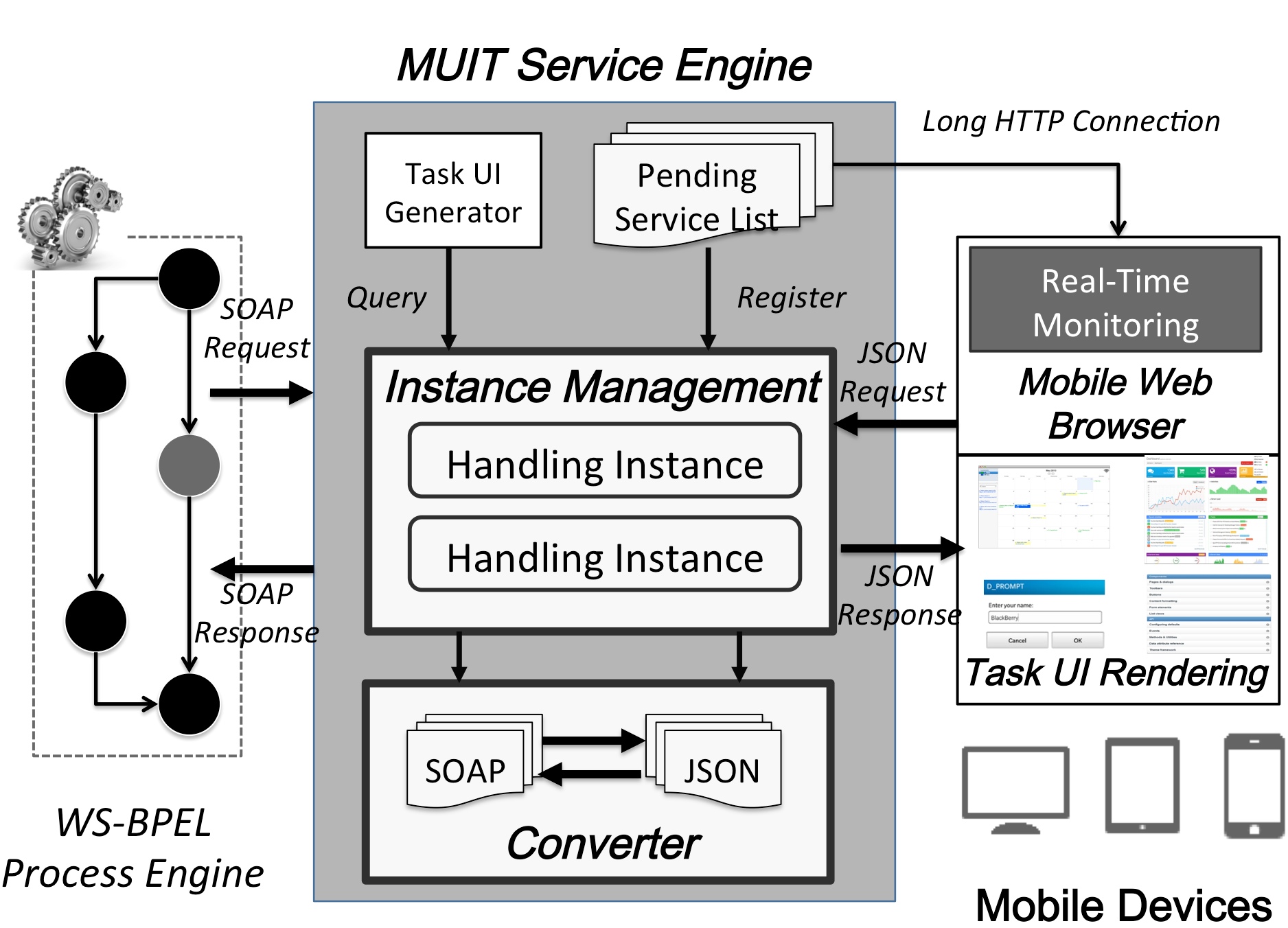}
\caption[7.5pt]{Interactions among BPEL Engine, MUIT Engine and Mobile Devices}\label{runtime}
\end{center}
\end{figure}

Based on the preceding design, this section describes the implementations and evaluations of MUIT prototype. Figure~\ref{runtime} illustrates how the MUIT engine communicates with WS-BPEL engine. In our current prototype, we deploy the MUIT engine on a commercial WS-BPEL compliant server, the Kingdee Apusic Platform Suite V6. Apusic Platform Suite provides standard infrastructural supports of Web services delivery, publish, WS-BPEL process modelling, and runtime engine. MUIT engine is deployed as a stand-alone component on the Apusic Platform, and can be accessed as a regular Web service. When a SOAP request arrives at the MUIT engine, Apusic Platform takes charge of processing underlying network communication protocols. When the user confirms the notification sent by the MUIT server and clicks through the hyperlink, he/she will be forwarded to his own mobile browser with the corresponding MUIT task UI. Since MUIT interprets the data model definition into regular Web pages, most currently popular browsers such as Safari, Chrome, and FireFox can execute MUIT task UI normally.

\indent We then provide some details on how to optimize runtime performance and the usability study from real-world scenarios.
\subsection{Communication Protocol Transformation}
\indent MUIT engine optimizes communication protocol between the mobile devices and the WS-BPEL. For each connected device, a \textit{Handling Instance} is created. Such \textit{Handling Instance} transforms SOAP request message into JSON format. Although SOAP messages are regarded as standard interoperability protocol for Web services, they are considered to be quite heavy-weight in terms of complex XML decoding and inefficient transmission performance~\cite{Pautasso:WWW08}. Since mobile devices are with relatively limited computation resources, JSON is then chosen as a promising option.

\indent After the message format is transformed to JSON, the \textit{Handling Instance} looks up corresponding MUIT tasks from the endpoint address and delivers it to the target mobile device. A new service request is put into the pending request queue. When the handling instance receives the response, it will transform JSON data into SOAP response and send it back to the WS-BPEL engine through the Apusic Server.

\indent To evaluate the performance, we use one popular device model, the LG Nexus 4,  as the testing device. Such a smartphone runs Android 4.2 OS, with a 1.5GHz Quad-core Snapdragon S4 Pro processor, 2GB RAM. We transfer the \textit{TaskApproval} data in SOAP format and HTTP-JSON format, respectively. We use the Android native \texttt{WebView} browser to access the MUIT engine under a Wi-Fi network. Then we compare the message size, parsing time, and energy. The message size and parsing time are simply extracted from the Android \texttt{proc} system file by looking up the \texttt{pid} of \texttt{WebView}. The energy consumption is computed based on the PowerTutor Android app~\footnote{PowerTutor. http://ziyang.eecs.umich.edu/projects/powertutor/}.

\begin{figure}
\centering
\begin{center}
\includegraphics[width=0.45\textwidth]{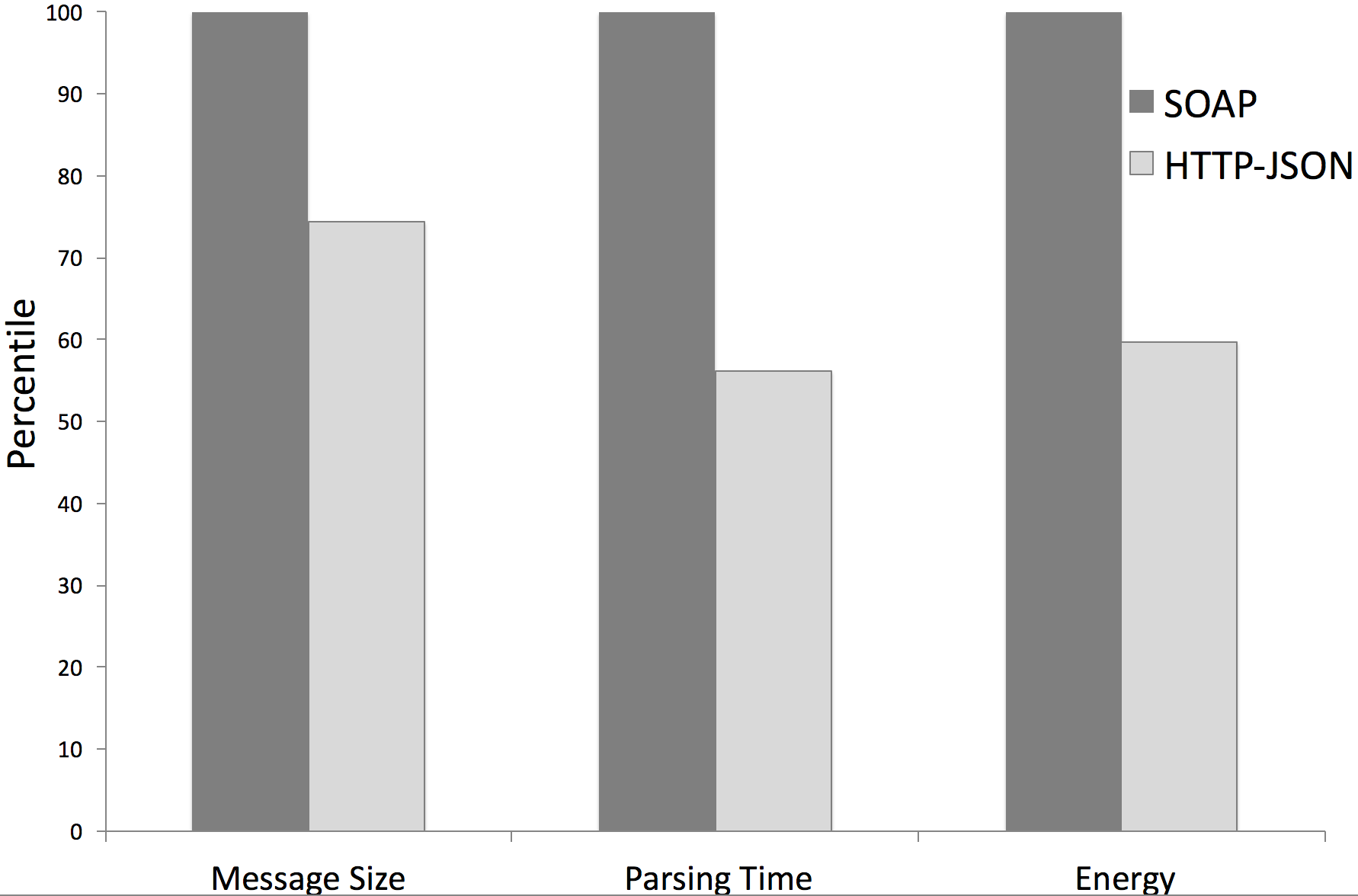}
\caption[7.5pt]{Communication Protocol Performance Optimization }\label{SOAPJSON}
\end{center}
\end{figure}

\indent Figure~\ref{SOAPJSON} illustrates the optimization contributed by our protocol optimization mechanisms. We use the performance of SOAP as a baseline (100\%). Obviously, the HTTP-JSON performs quite better with 25.5\% message size reduced. The reason is that HTTP-JSON does not require SOAP-Envelope header information, but only a plain string instead. Correspondingly, mobile computation resources are saved without decoding SOAP XML messages. The parsing time and energy are saved by 44\% and 41\%, respectively.  
\subsection{Concurrent Devices Management}
On the MUIT engine, the \textit{Handling Instance} takes charge of dealing with several concurrent requests from a number of mobile devices. As people may not always deal with tasks in time, pending requests have to wait for a long time before the task is completed. However, if there are too many pending service requests, the MUIT engine will probably become slow or even overloaded. To deal with this problem, we provide a passivation mechanism to suspend these unfinished service instances until human tasks are completed. 

\indent We deal with \textbf{asynchronous} and \textbf{synchronous} requests in different manners. When passivating an asynchronous request, the MUIT engine serializes the instance state, stores the state with the instance's callback address, and releases the instance. When the user response from the \textit{WS-BPEL} engine is received, the passivation mechanism  restores the instance from the storage, and sends the results back to the mobile device. Passivating the synchronous requests is quite different, since these requests need to hold live connections with the WS-BPEL engine. When passivating a synchronous request, our mechanism needs to make the instance sleep, and awakes it when the response returns.

\begin{figure}
\centering
\begin{center}
\includegraphics[width=0.45\textwidth]{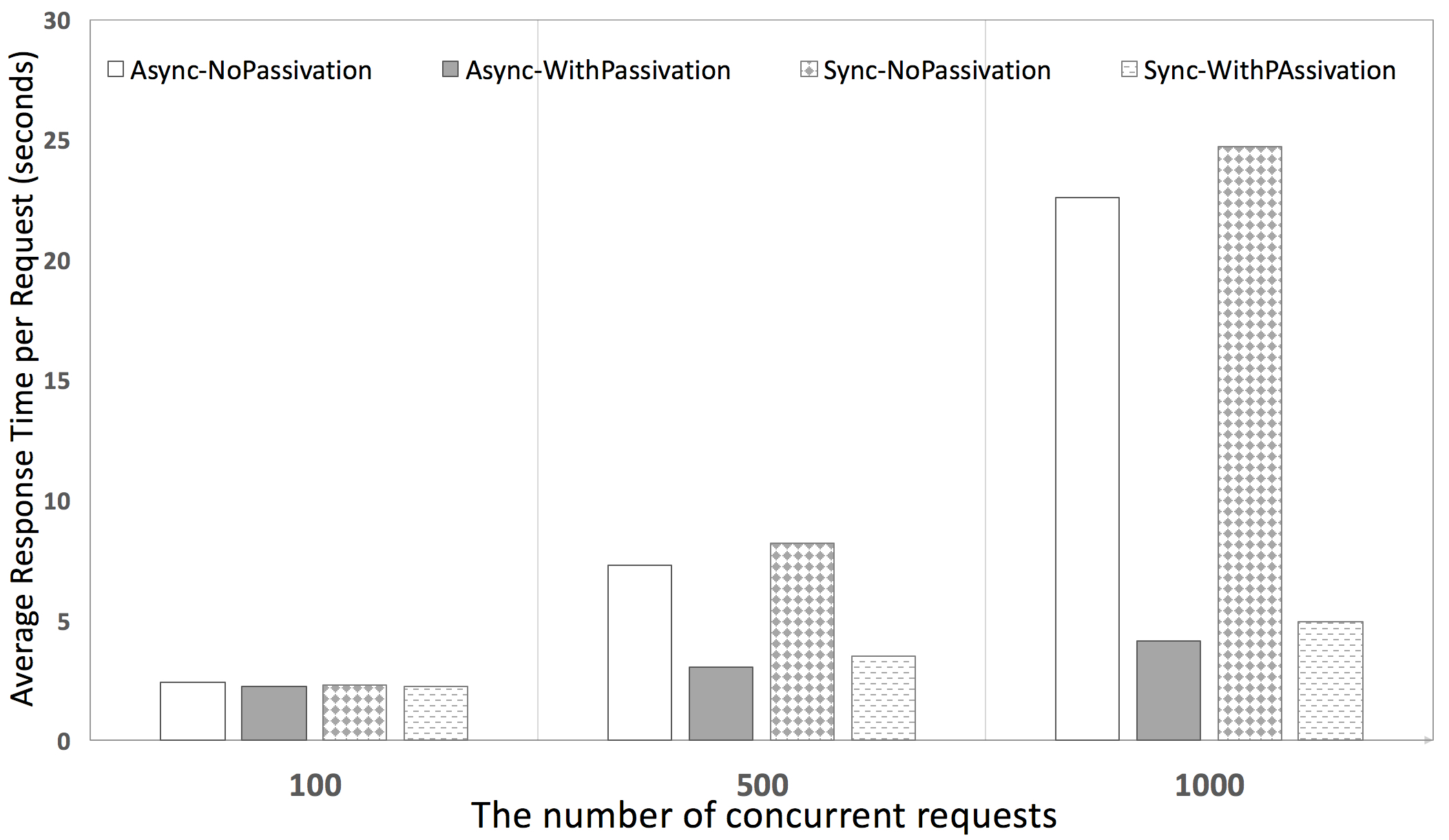}
\caption[7.5pt]{MUIT Service Instance Management}\label{passivation-time}
\end{center}
\end{figure}

To demonstrate the effect of our passivation mechanism, we make a simulation testing for concurrent request management. We suppose that each task request is successfully processed in 2 seconds. We then set the number of concurrent requests of 100, 500, and 1,000, respectively. Under each scenario, we assume that 20\% requests need to wait for human processing for 1 minute. We then validate the \textit{average response time} (abbreviated as ``\textit{ART}")  for other 60\% requests. We test both asynchronous and synchronous requests with/without enabling the passivation mechanism, respectively. The maximum size of the threads pool on the MUIT server is assigned as 1,000. The results are reported in Figure~\ref{passivation-time}. 

Not surprisingly, the \textit{ART} degrades with the increasing number of concurrent requests, as the number of available resources on server also reduces. However, it is observed that the passivation mechanism improves the \textit{ART} significantly. By ``freezing" the currently inactive request,  server-side resources are saved for serving other active 60\% requests, and the \textit{ART} increases slightly. For example, for the asynchronous requests, the passivation mechanism can reach the \textit{ART} of 2.25 seconds with 80 concurrent requests, but reach up to only 4.01 seconds with 800 concurrent requests. In contrast, without enabling passivation, the \textit{ART} increases from 2.4 seconds to 22.6 seconds. It implies that our mechanism can be scalable to the varying number of concurrent requests.

Additionally, with our passivation mechanism, the synchronous request usually accounts for more \textit{ART} than the asynchronous request. It is not surprising because MUIT needs to maintain additional network resources to keep live connections for synchronous requests. 
\subsection{Context-Aware Optimization}
\indent Due to human mobility, the network connections are likely to be unreliable or even unavailable. Then, the results made on mobile devices cannot be submitted to MUIT engine in time. More seriously, sometimes the timeout exception can lead to errors. To overcome the unreliable network connection, we evaluate the offline cache mechanism in MUIT. The offline cache is allowed to be triggered when the network condition changes. The data and states are maintained locally at device side. When the network connection recovers, the results can be synchronized to MUIT engine and then forwarded to the WS-BPEL server. Certainly, the corresponding instance on MUIT engine needs to maintain its state and assure the consistency.  

\indent The offline cache can also be employed when people are interrupted for a moment. In this way, the offline cache mechanism is triggered by a timeout (e.g., 60 seconds without interaction). Then, we close the connections between the mobile devices and the MUIT engine server. 

\begin{figure}
\centering
\begin{center}\subfigure[User Clicks\label{click}]{
\includegraphics[width=0.45\textwidth]{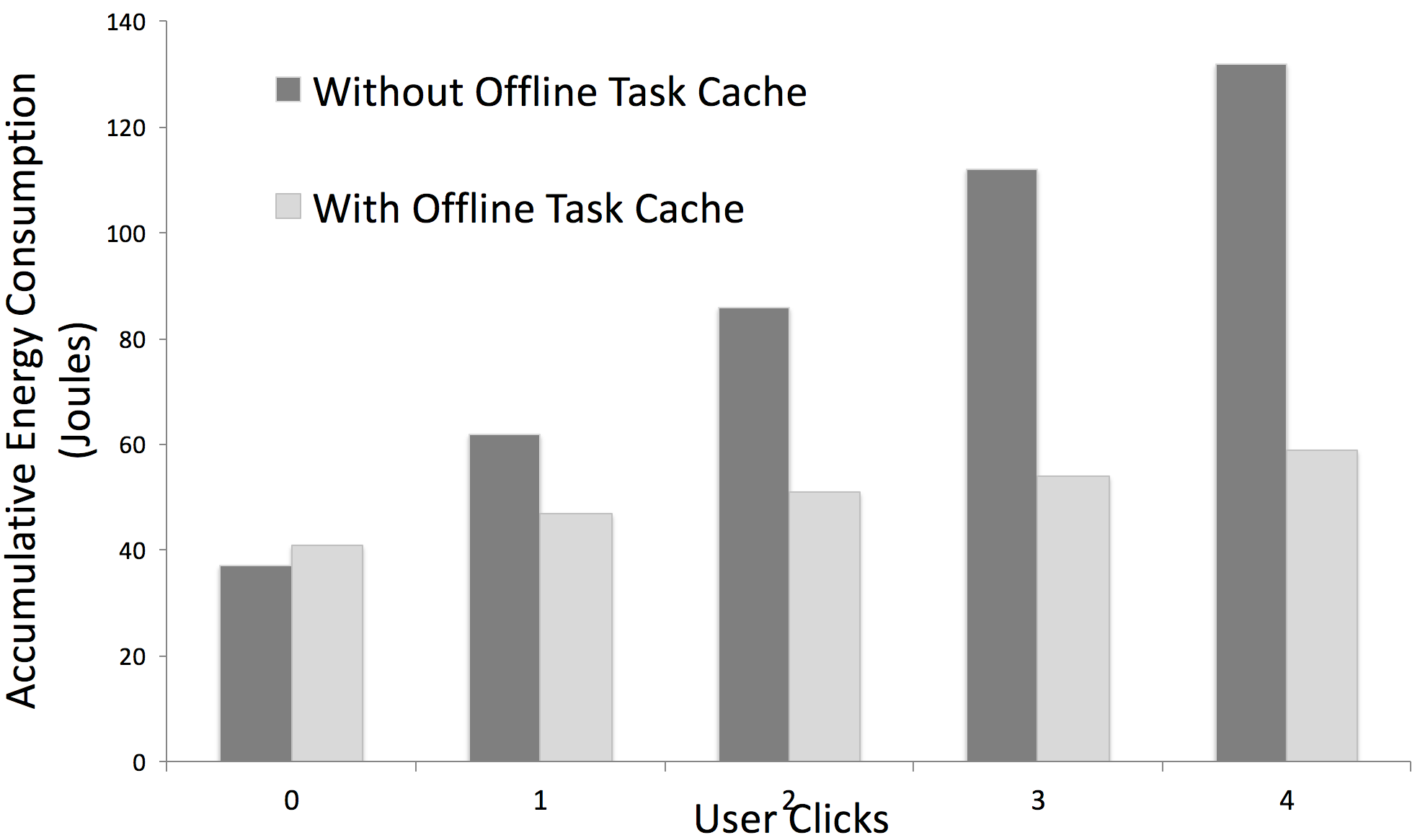}}
\subfigure[Session Time\label{session}]{
\includegraphics[width=0.45\textwidth]{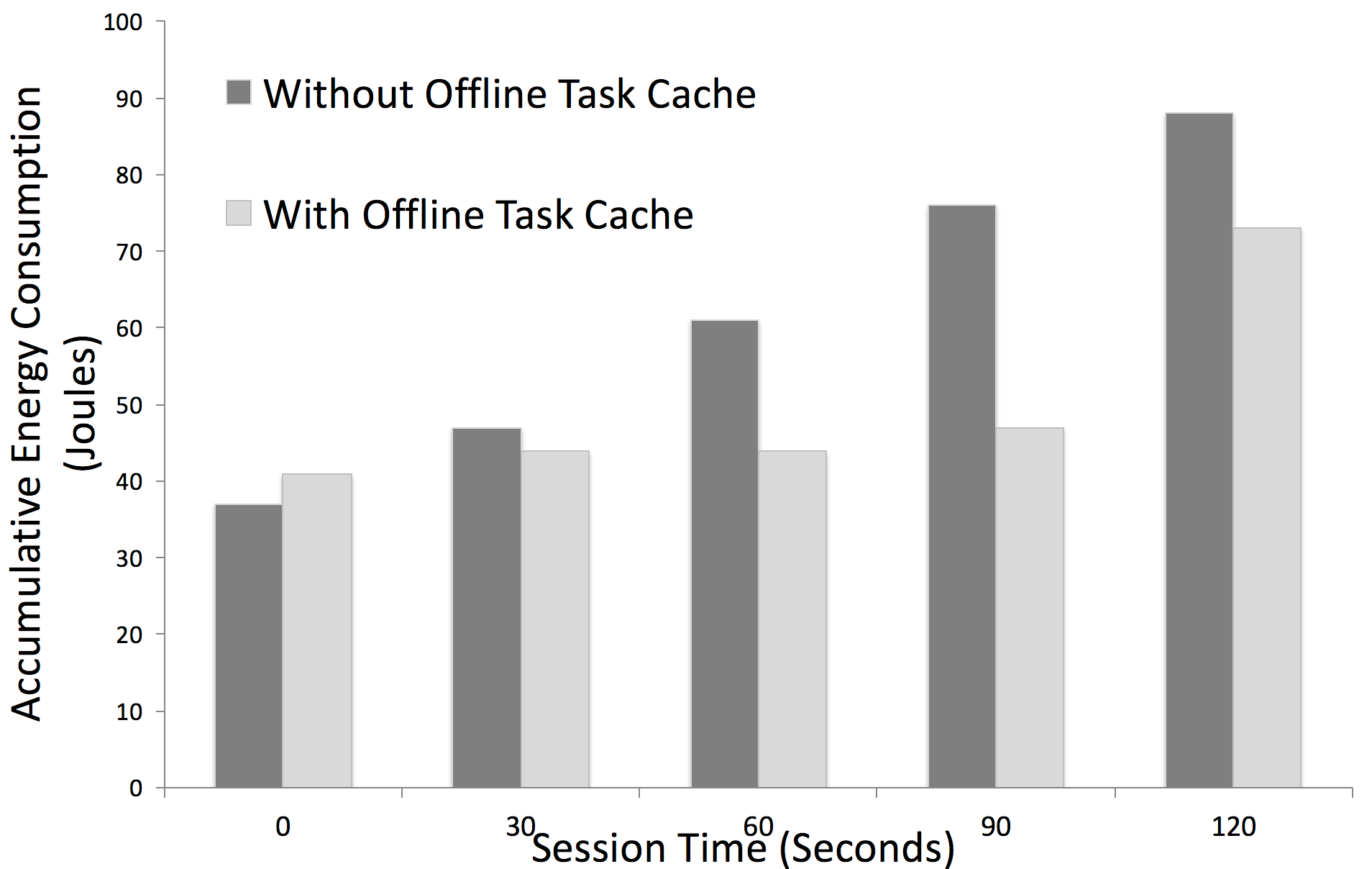}}
\caption[7.5pt]{Device Energy Optimization by Offline Cache}
\end{center}
\end{figure}

\indent We then evaluate the performance optimization gained by offline cache. We use the \textit{TaskApproval} UI as an illustrating example. We import the UI code with and without offline cache, respectively. Then we investigate how offline cache can optimize energy consumption, by summarizing the \textit{accumulative} energy consumption in Figure~\ref{click}. First, we observe that the \texttt{With\_Cache} consumes more energy than the \texttt{Without\_Cache}, when the page loads (click 0). It is not surprising as the browser needs to interpret offline cache logics and store data. Next, we click the search query button in Figure~\ref{search} from 1 to 4 times. For each click, the \texttt{With\_Cache} performs only the search over local storage, and can reduce energy consumption by avoiding network connections.

\indent A similar experiment is conducted by setting different session length. In this experiment, we assume that the device is idle for 2 minutes. The timeout is set to be 30 seconds when no user interactions are detected. The \texttt{With\_Cache} mechanism is triggered to freeze connections and no energy is consumed after 30 seconds. In contrast, the \texttt{Without\_Cache} keeps network connections all the time. However, some additional but unnecessary energy is consumed. At the end of 2 minutes, we recover the network connections.

\indent The experiments implies that, although offline cache may introduce slight code parsing overhead of its first load, it is still worth employing such mechanism to improve device energy consumption.
\subsection{Usability Study}
Finally, we make a filed study to evaluate the usability of MUIT with several scenarios. The MUIT engine has been deployed with Apusic Platform in four typical scenarios: \textbf{logistic, sales, healthcare}, and \textbf{travel}. Common user interfaces in these scenarios include: \emph{ (1) fill in an electronic form or document; (2) task status query; (3) notification and alert; (4) modify task information}. 

The field study is composed by 24 end-users who used to participate in some business processes on desktop PCs. Each scenario is assigned with 6 individual users. Each user is equipped with a mobile device. To test the cross-platform portability, the selected devices include 4 smartphones (iPhone 5s, Samsung Galaxy Note 2, Nexus 4, and Windows Lumia 520), and two tablet computers (iPad 2, Samsung Galaxy Tab3). 

\indent In terms of user-related metrics to be evaluated, we first demonstrate all features provided by MUIT. We allow users to trial MUIT under the supervision of an evaluator guiding them in performing the assigned tasks. After the training stage, the tasks are executed in the field. We control the network connection (available and unavailable) in the experiment. The system successfully adapts to the changes. Each task execution is followed by structured interviews, where the users are asked to fill in an questionnaire based on Linkert scale~\footnote{Linkert scale: http://en.wikipedia.org/wiki/Likert\_scale is a common user survey metric}. Scores ranged from 1-5 (\textit{strongly unsatisfying, unsatisfying, neutral, satisfying, and strongly satisfying}) to present user satisfactions. Users are allowed to present their reviews or suggestions to each question. 

We totally collect 87 independent questionnaires and summarize the questions to map MUIT features. For example, the feature ``\emph{Display and Font}" in the questionnaire corresponds to the question``\emph{Required items all normally displayed on my device screen}" and ``\emph{Fonts and pictures look comfortable}". Then we compute the average score for each feature to evaluate user experiences. The results are shown in Table~\ref{usability}. 

\indent We find that the  features such as \textit{Sub-Page Navigation}, \textit{Notification}, and \textit{Offline Operation}, cater for the mobile users' requirements quite well. Especially, users from sales claim that ``\textit{Offline Operation}" feature is extremely useful by commenting ``\emph{It allows me to sign contract with customer anywhere}".

\begin{table*}\small
\renewcommand{\arraystretch}{1.3}
\caption{MUIT Usability Evaluation}
\label{usability} \centering
\begin{tabular}{|l|c|c|c|c|}
\hline
\backslashbox{Features Score}{Scenario (Tasks)} & \bfseries Sale (20) &  \bfseries Logistic (22) & \bfseries HealthCare (19) & \bfseries Travel (26)\\
\hline
\textbf{Display and Font}& 4.3 & 4.6 & 3.9 & 4.7\\
\hline
\textbf{Touch Controls and Guestures}& 4.0 & 4.4 & 3.7 & 4.5\\
\hline
\textbf{Sub-Page Navigation}& 4.6 & 4.6 & 4.4 & 4.7\\
\hline
\textbf{Notification}& 4.8 & 4.7 & 4.8 & 4.8\\
\hline
\textbf{Responsiveness}& 4.2 & 4.1 & 4.7 & 4.4\\
\hline
\textbf{Integrated Widgets}& 3.7 & 4.1 & 4.2 & 4.4\\
\hline
\textbf{Offline Operation}& 4.8 & 4.8 & 4.7 & 4.8\\
\hline
\end{tabular}
\end{table*}
\indent It can be found that the scores vary slightly among different domains. For example, the feature ``\textit{Integrated Widgets}" is not highly ranked by the users from sales. They argue that ``\emph{Calendar and email are good, but it would be better if I can share with my friend via WeChat or Weibo~\footnote{WeChat and Weibo are two popular social networking apps in China. WeChat provides the similar functionalities such as WhatsAPP while Weibo is similar to Twitter}}". It indicates that some potential exploration of MUIT should be considered. We learn that users prefer integrating more smartphone apps by MUIT. 

Some users from healthcare complain that \textit{Touch controls} of MUIT are not feasible enough when they process electronic patient records. A possible reason is that, typing words with virtual \texttt{QWERTY} keyboard is not proper for continuously editing long documents. Some more advanced inputs beyond touch controls, such as Speech-to-Text should be considered to be integrated in MUIT in the future. 

\indent We also find that, the screen estate exactly impacts on some features such as \emph{responsiveness}. Some tablet users comment that ``\emph{Adaptation of portrait or landscape is interesting, but I am also OK without it}". However, most smartphone users click this feature as ``\textit{strongly satisfying}". 

\indent From user feedbacks, we can confirm that MUIT is useful and effective in most WS-BPEL scenarios involved in our field study. 

\section{Related Work}
In previous sections, we illustrated the MUIT design and implementation, and evaluated the performance and effectiveness. In this section, we discuss the related efforts in literature from two aspects, i.e., \textit{user interfaces modeling in workflow} and \textit{domain-specific languages for Web applications}.
\subsection{User Interfaces Modeling in Workflow}
Integrating human intervention into workflows is an important aspect in the research body of service-oriented business process management~\cite{Tan:TASE09}. Not limited to WS-BPEL, a number of efforts have been made to enable human tasks in workflows, by proposing some new descriptions or extensions beyond existing process specifications~\cite{Montagut:CollabCom05}\cite{Kunze:JCP07}\cite{Dustdar:IC08}\cite{Dustdar:IC10}. As mentioned previously, WS-BPEL is the dominant standard for service-oriented business process. But it was originally proposed to enable automated Web services orchestration~\cite{Tan:TASE10} without considering human interactions. To address such limitations, two extensions, BPEL4People~\cite{IBM:WS4People07} and WS-HumanTask~\cite{IBM:HumanTask07}, were proposed. However, in practice, these two extensions are not ever successful as expected. The main problem of BPEL4People is that the \textbf{\emph{PeopleActivity}} cannot be directly and seamlessly integrated into existing WS-BPEL engine. It needs to import a new element such as $<$\textbf{\emph{b4p:peopleActivity}}$>$ and inevitably instruments current WS-BPEL infrastructures, thereby introducing additional maintenance cost~\cite{Qiu:ICWS08}. Another reason is that, the two extensions mostly concentrate on the specification of manual human-oriented tasks, but lack the possibility to describe a user interface in detail~\cite{Zaplata:VBK09}~\cite{Gerardo:PESOS09}. 

\indent In practice, human tasks usually require user experiences such as data visualization, form, dashboard, and so on. Gerardo \textit{et al}~\cite{Gerardo:PESOS09} proposed that Web applications were considered to be a promising solution for SOA application presentation. CRUISE was then designed for building rich web UI for WS-HumanTask~\cite{Stefan:WISE09}. Distributed UI orchestration designed a mashup-like and process-based approach to aiding developers and implementing UI orchestrations~ \cite{Florian:InfoSys12}. Similar to CRUISE, they equipped the \textit{WSDL4UI} and \textit{BPEL4UI} model with WSDL and BPEL extensions. Our previous iMashup platform was also applied in WS-BPEL process on desktop web browsers~\cite{Zhao:ICWS10}. These efforts provide primitive experiences of Web technologies into business process management systems.

\indent Assembling mobile devices into enterprise business process is not entirely new. A lot of efforts have been proposed. Essentially, supporting distributed orchestration of UI in workflows should address technical issues such as decentralized enactment~\cite{Pantazoglou:TSC14}, cross-realm security~\cite{Xu:TSC12}, and process reliability~\cite{Zheng:TC14}, etc. \textbf{It should be emphasized that, this paper's focus is not orchestrating distributed mobile devices, but developing, realizing, and integrating user interfaces for human tasks in WS-BPEL}. Actually, an architecture supporting distributed execution of workflows in pervasive environments needs to be conducted~\cite{Montagut:CollabCom05}. Allocating tasks across mobile devices and orchestrating them will be considered in our future work. Hence, in this paper, we only focus on how to support user interfaces on mobile devices in workflows. 

Hackman \textit{et al.}~\cite{Hackman:ICSOC06} firstly proposed a conceptual proof of involving feature phones. Pryss \textit{et al.}~\cite{Pryss:Caise10} proposed MARPLE as a tight integration of process management technology with mobile computing frameworks in order to enable mobile process support. Russo \textit{et al.}~\cite{Russo:SPE12} proposed ROME4EU, a mobile process-aware information system that was developed for the coordination of emergency operators. Zaplata \textit{et al.}~\cite{Zaplata:BPM09}\cite{Zaplata:VBK09} proposed an abstract user interface model that can be applied on J2ME-compliant mobile devices such as PDAs. In addition, they suggested that thin client approaches like Web technologies were limited in terms of unreliable network and rich user experiences support on PDAs. 

\indent Compared to previous efforts, MUIT differs in terms of various unique features. MUIT targets at modern popular smartphones and tablet computers. MUIT adopts Web UI that not only realizes the platform-neutrality, but also provide high-quality user experiences. To support the mobile-specific features, MUIT carefully designs a domain-specific language to aid developers in avoiding programming complexity and mixed code. The performance and usability of MUIT are evaluated with a commodity WS-BPEL engine. 
\subsection{DSL for Web applications}
Due to the complex and trivial programming efforts of Web applications, some DSLs are proposed. Visser \textit{et al.} proposed WebDSL~\cite{visser2008:webdsl}, a well-known DSL for the development of Web applications. WebDSL can generate Java Web applications that can be deployed in Java servlet container. To accommodate the mobile devices, WebDSL is further extended to Mobl~\cite{Hemel:OOPSLA11}. Mobl developed some built-ins such as page and screen, which can encapsulate the basic display units on mobile devices. Indeed, MUIT learns the successful experiences and inherits some similar constructs from preceding DSLs. However, MUIT is further \textit{domain-specific}  by advancing itself for the mobility-support in existing WS-BPEL infrastructures. First, MUIT does not introduce new design patterns but enhances the traditional Model-View-Controller by alleviating the complexity from repeatedly similar code. Second, the DSL compiler of MUIT pre-processes the WSDL of a Web service and generates an intermediate structure of UI. Such feature can relax the developers from tedious efforts on dealing with creating a new UI. Third but not the last, the features of MUIT are particularly designed for adapting to the various environments in mobile workflow, such as the push-oriented notification, the possible network failures and different OSes, and the changes of user interaction. For example, MUIT allows to cache the state of mobile Web applications other than data only, and reduces the unnecessary network and energy consumption. Such features are quite useful from the survey of user feedbacks.   
\section{Discussion}
Although the evaluations can demonstrate the effectiveness of MUIT, some limitations and todo issues should be discussed, which are significant to reach more coverage and adoption of MUIT.

Currently, MUIT adopts the Web applications to reach the cross-platform portability over iOS, Android, BlackBerry, and so on. It is commonly assumed that the Web applications is usually worse than that of native apps, which can be a potential obstacle to applying MUIT in more contexts. However, our recent study~\cite{Liu:ICWS2015} finds some counter-intuition observations that the preceding assumption could not be always true, i.e., the Web apps can win the native apps in terms of response time, traffic volume, and energy. Furthermore, our experiences work~\cite{Ma:WWW2015} also evidences that the performance of mobile Web apps can be further optimized, by caching both the application and the data locally at finer granularity. 

From our survey, it is reported that the users would like to use the device services such as GPS, camera, bluetooth, and so on, in their activities of business process. For example, someone argue that ``\textit{I prefer to taking photo of the e-proof of receipts and upload as soon as possible, but MUIT doesn't support.} " Indeed, the Web apps have limitations to access some device services. HTML5 offers many JavaScript APIs that give access to various device services, but their implementation in mobile devices is not always complete. Access to audio and video services is limited — it is possible to play an audio or video file, but only by launching the dedicated audio or video player. Access to other device-specific features such as bluetooth, the built-in compass, camera and local file storage are not supported yet. Some recent efforts can be considered and further leveraged, e.g., using WebSockets and HTTP for efficient bi-directional communication between Web apps and the device-specific service~\cite{Puder:MSES14}. Another option is to employ some frameworks that can encapsulate and deploy the Web apps in form of hybrid apps according to the target OSes, e.g., the Apache Cordova~\cite{apachecordova}. However, it is highly debatable the additional performance overhead and even crash-down threats to MUIT users. We plan to address the device-local support in our future work.

\section{Conclusion and Future Outlook}
Addressing the mobility and user interactions in WS-BPEL process, we presented the \textbf{MUIT} middleware, which designs a programming abstraction with a domain-specific language, realizes the adaptive mobile Web user interfaces, and seamlessly integrates the Web UI into standard WS-BPEL. We evaluate MUIT in terms of performance and usability.

\indent Although MUIT is currently designed and implemented as an add-on for WS-BPEL engines, we think that it can be extended to other popular workflows such as WSCI and PDDL by adapting its \textit{Service Interface} to communicate with the components in corresponding platforms. 

\indent  One ongoing effort is developing supporting tools to reduce programmers learning curve on using MUIT. Another progress is to explore browser-kernel-level extensions to enable data exchange and communication between MUIT Web applications with other native apps. Supporting orchestration of these distributed mobile users and their screens~\cite{Pantazoglou:TSC14} and avoiding potential cross-organization security~\cite{Badr:TSC12} are also worth investigating to meet the \textbf{BYOD} requirements for enterprise.

\bibliographystyle{IEEEtran}
\bibliography{IEEEabrv,bibtex}





%
%
%
\begin{IEEEbiography}[{\includegraphics[width=1in,height=1.25in,clip,keepaspectratio]{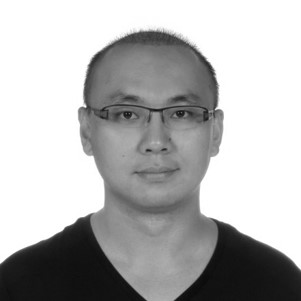}}]{Xuanzhe~Liu}
is an associate professor in the School of Electronics Engineering
and Computer Science, Peking University, Beijing, China. His research interests
are in the area of services computing, mobile computing, web-based systems, and big data analytic. For more information, please visit his webpage via (http://www.sei.pku.edu.cn/~liuxzh).
\end{IEEEbiography}
\begin{IEEEbiography}[{\includegraphics[width=1in,height=1.25in,clip,keepaspectratio]{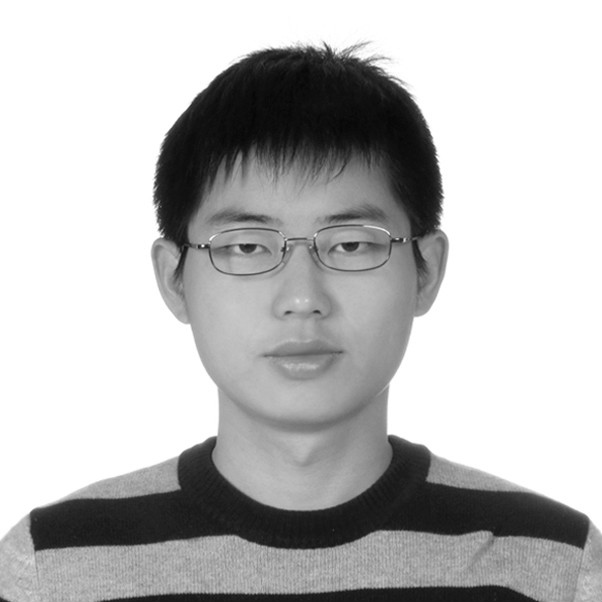}}]{Mengwei~Xu}
is now a senior student in the School of Electronics Engineering and Computer Science of Peking University, Beijing, China. His research interests include service computing and mobile computing.
\end{IEEEbiography}
\begin{IEEEbiography}[{\includegraphics[width=1in,height=1.5in,clip,keepaspectratio]{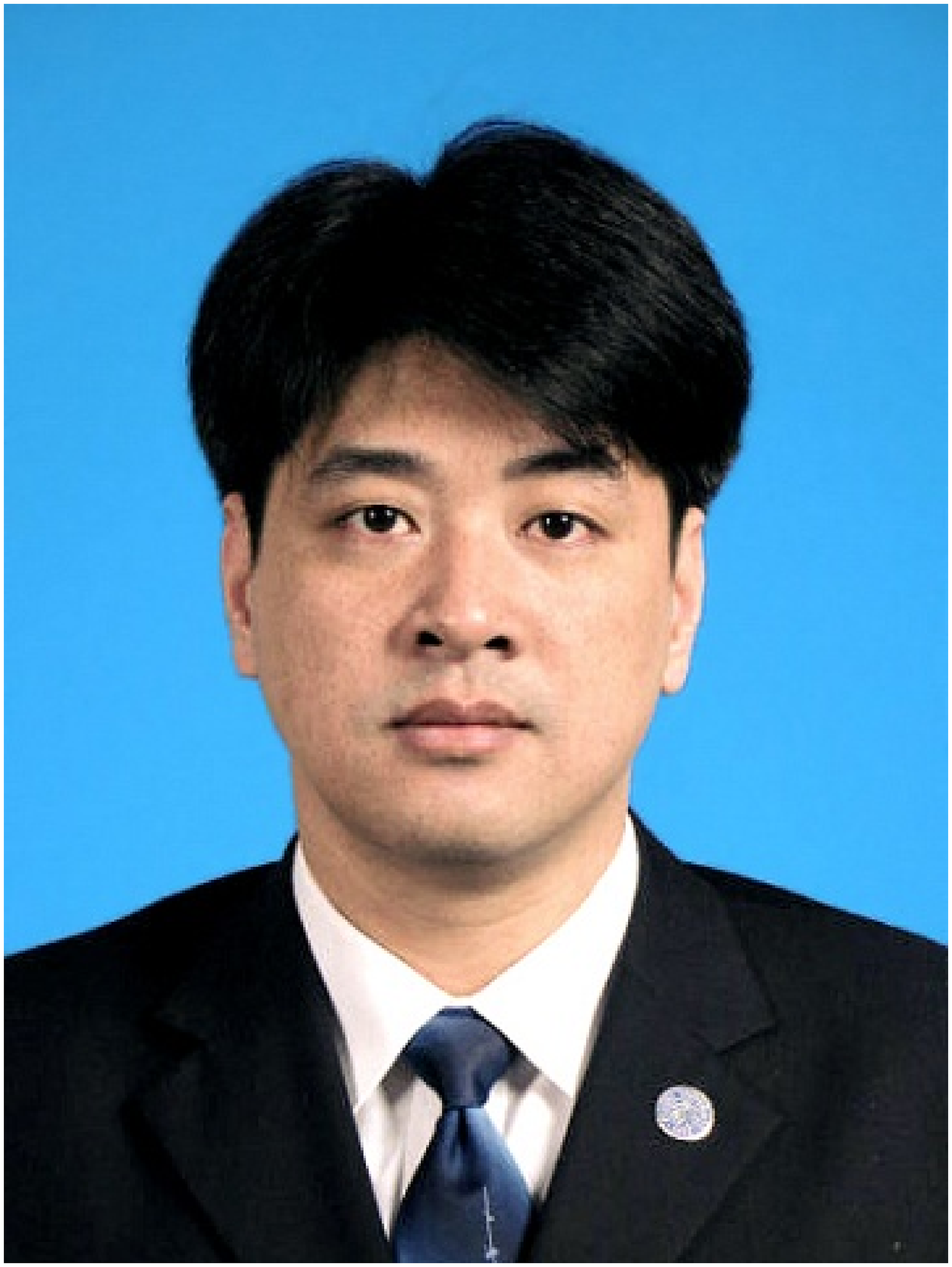}}]{Teng~Teng}
is now the director of Middleware Research Lab, Kingdee Ltd. His research interests include middleware, service computing and business process management.
\end{IEEEbiography}
\begin{IEEEbiography}[{\includegraphics[width=1in,height=1.25in,clip,keepaspectratio]{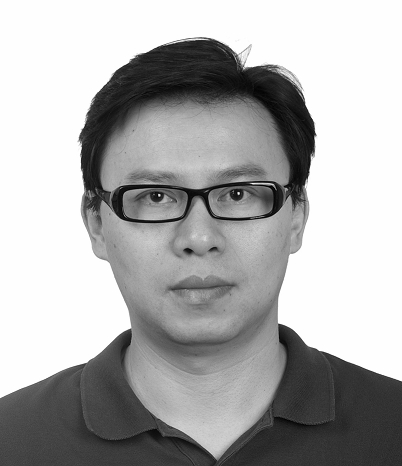}}]{Gang~Huang}
received his Ph.D degree in the School of Electronics Engineering and Science of Peking University, Beijing, China. He is now a full professor in Institute of Software, Peking University. He is research interests are in the area of middleware of services computing and mobile computing. He is a member of IEEE.
\end{IEEEbiography}


\begin{IEEEbiography}[{\includegraphics[width=1.3in,height=1.25in,clip,keepaspectratio]{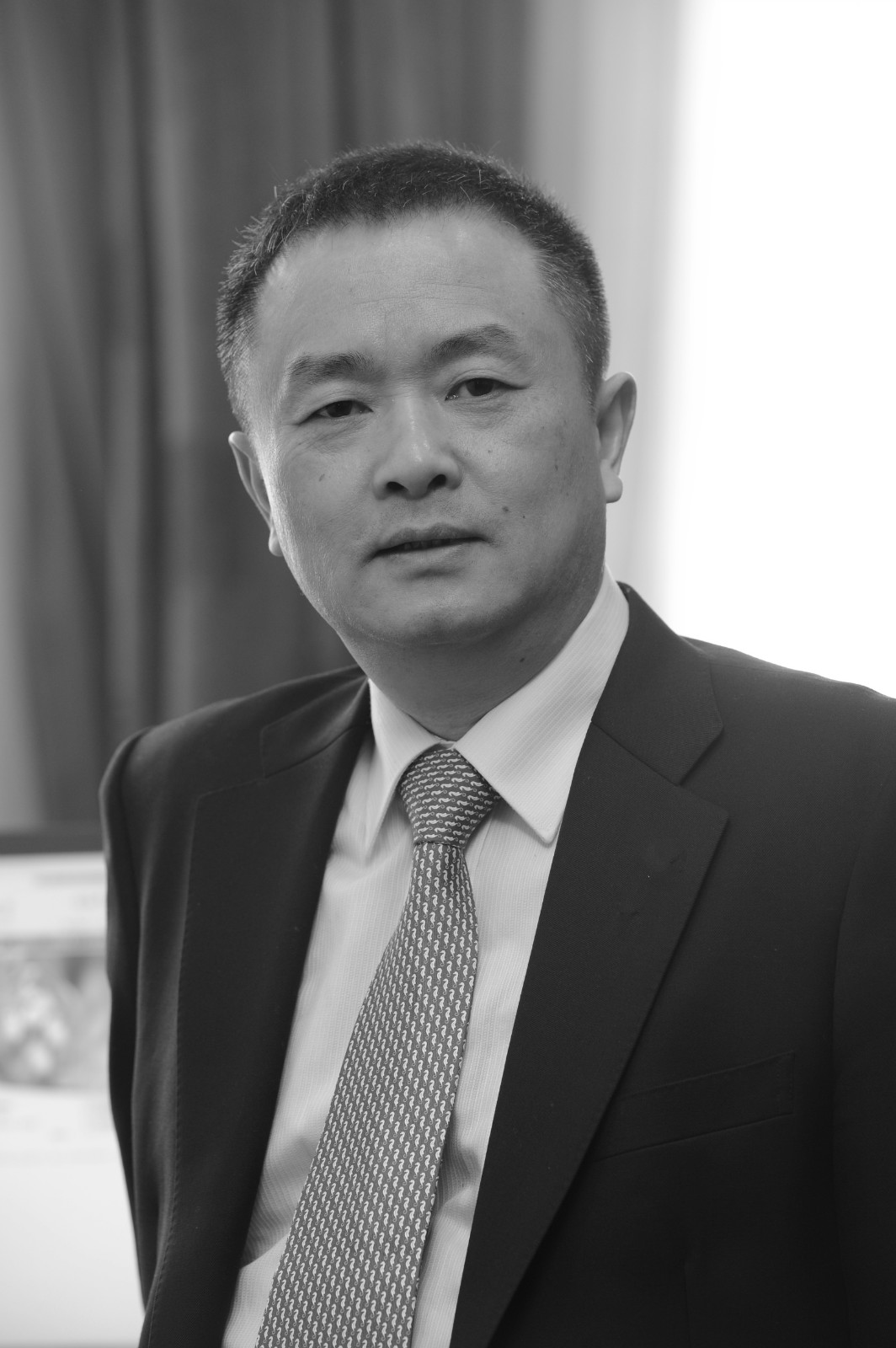}}]{Hong~Mei}
is a full professor of School of Electronics Engineering and
Computer Science, Peking University, Beijing, China. His current research interests are in the area of software engineering and operating systems. He is the Member of Chinese Academy of Sciences, and the Fellow of China Computer Federation (CCF). He is a fellow of the IEEE.
\end{IEEEbiography}
\balance

\end{document}